\documentstyle{article}
\begin{document}

\title{Renormalization of the Lattice HQET Isgur-Wise Function}

\author{Joseph Christensen\footnote{Present address: McMurry University,
Abilene, TX 79697},
Terrence Draper \\
{\it University of Kentucky, Lexington, KY 40506}
\bigskip \\
and Craig McNeile\\
{\it University of Liverpool, Liverpool, L69 3BX}}

\date{}

\maketitle

\begin{abstract}

We compute the perturbative renormalization factors required to match to the
continuum Isgur-Wise function, calculated using lattice Heavy Quark Effective
Theory.  The velocity, mass, wavefunction and current renormalizations are
calculated for both the forward difference and backward difference actions for
a variety of velocities.  Subtleties are clarified regarding tadpole
improvement, regulating divergences, and variations of techniques used in these
renormalizations.

\end{abstract}

\bigskip\bigskip

\centerline{PACS numbers: 12.38.Gc,12.39.Hg}

\vspace{-4.5in}
\hfill \begin{tabular}{l} UK/99-22\\December 1999\end{tabular}

\vfill

\pagestyle{headings}
\renewcommand{\thesection}{\Roman{section}}
\renewcommand{\thesubsection}{\Roman{section}.\Alph{subsection}}

\newpage

\section{Introduction} \label{c:intro}

The unitarity of the Cabibbo-Kobayashi-Maskawa (CKM) matrix is regarded as a
crucial test of the standard model~\cite{Neubert:1996qg,Rosner:1998hz}; the
precise determination of these matrix elements has received extensive
experimental and theoretical scrutiny.  The $V_{cb}$ CKM matrix element can be
extracted from the reaction $B \! \rightarrow \! D^\ast \, l \,
\overline{\nu}_l$, if the theoretical factors in the decay rate can be reliably
computed. The heavy quark effective field theory (HQET) formalism is
well-suited to the analysis of this decay.  The differential decay rate of the
above process is
\begin{equation}
\label{eq:decay rate}
\frac{d}{d(v \cdot v')} \Gamma(B \rightarrow D^* \, l \, \overline{\nu}_l)
 = \frac{G^2_F}{48 \pi^2} k(m_B, m_D, v\cdot v') \left| V_{cb} \right|^2
\xi^2(v \cdot v')
\end{equation}
where $\xi\!\left(v \cdot v'\right)$ is a universal form factor, the Isgur-Wise
function.  The function $k$ can be calculated in perturbation theory using
various approximations~\cite{Neubert:1996qg,Czarnecki:1997wy}.  The Isgur-Wise
function is a QCD matrix element that must be computed non-perturbatively.
Previously and in a companion paper~\cite{Christensen97b,Christensen98}, we
discuss the numerical calculation of the Isgur-Wise function using lattice
HQET\@.  In this paper, we discuss the perturbative matching of lattice HQET to
continuum HQET, which allows the conversion of the results from the numerical
simulations into physical predictions.  Specifically, we shall be matching from
the lattice to the continuum matrix element,
\begin{equation} \label{eq:current}
\left< D,v \left| J^{b \rightarrow c}_\mu\!\left(0\right) \right| B,v'  \right>
 = \sqrt{ M_D M_B } \left( v_\mu + v'_\mu \right) \mbox{$\xi( v \cdot v' )$}
\end{equation}
where $v$ and $v'$ denote the 4-velocities of the $c$ and $b$ quarks, and
\begin{equation}
J^{b \rightarrow c}_\mu\!\left(x\right)
 = \bar{c}(x) \gamma_\mu \left( 1 - \gamma_5 \right) b(x)
\end{equation}
is the weak current for the transition of a bottom to a charm
quark~\cite{Isgur8990}.

The Isgur-Wise form factor describes the response of the quark-gluon sea
surrounding the heavy quark due to a sudden change in velocity of the heavy
quark when it decays.  In HQET, the Isgur-Wise function is non-perturbatively
equal to one at the point of zero recoil, $v=v'$; HQET does not constrain the
Isgur-Wise function at non-zero recoil.  Continuum perturbative corrections are
required to obtain the zero recoil result in QCD; however, these are known to 2
loops~\cite{Czarnecki:1997wy}.  Unfortunately, there is no experimental data at
zero recoil, so the experimental data is extrapolated~\cite{Neubert91} to zero
recoil in order to estimate $V_{cb}$ using Eq.~(\ref{eq:decay rate}).
Knowledge of the functional form of the Isgur-Wise function would greatly aid
this extrapolation.  The Isgur-Wise function can be calculated
non-perturbatively, in principle, from QCD for arbitrary recoil.  In our
companion paper~\cite{Christensen98}, we describe our simulations that use
lattice HQET to calculate the Isgur-Wise function.

There have been previous calculations of the renormalization factors for
lattice HQET\@. Unfortunately, not all of the perturbative factors required for
our numerical simulations were calculated.  After lattice HQET was introduced
by Mandula and Ogilvie~\cite{Mandula92}, there were a number of concerns about
the validity of the lattice HQET
formalism~\cite{Aglietti:1992ta,Aglietti:1992in,Sachrajda:1993hu}.  The
consistency of lattice HQET was finally demonstrated by
Aglietti~\cite{Aglietti94} in perturbation theory.  However, Aglietti used a
form of lattice HQET action that is less convenient for numerical simulation
than the one originally used by Mandula and Ogilvie.  The difference between
the HQET actions was in the use of a forward or backward finite difference in
the time direction (see Sec.~\ref{s:lattice}).  Also, Aglietti considered only
a special kinematic limit with one quark at rest and the other quark at finite
velocity.  Mandula and Ogilvie~\cite{Mandula:1997hb} limited their work to the
velocity renormalization factors for the forward-difference action (which we
used in our simulations); they calculated neither the vertex function nor the
wavefunction renormalization which are required to renormalize the lattice
data.

In this paper, we calculate the perturbative factors required to renormalize
the Isgur-Wise function obtained from a lattice HQET simulation.  The
calculation includes two HQET actions: one with the forward time derivative and
one with the backward time derivative.  We follow the formalism developed by
Aglietti~\cite{Aglietti94}, but generalize Aglietti's expression for the vertex
function to arbitrary input and output velocities (as is required for the
analysis of the simulation data).  We also include the effects of tadpole
improvement and discuss a subtlety in the calculation of the vertex function.

Section~\ref{s:continuum} will provide a sparse review of continuum HQET in
order to put the lattice calculation into context.
Section~\ref{s:lattice} will describe the details of the velocity, mass,
wavefunction, and vertex renormalizations for the lattice actions, including a
discussion of the ``reduced'' results and an evaluation at non-zero recoil.
Section~\ref{s:matching} will describe how these are combined into a single
renormalization for the lattice current to be matched to the continuum.
Section~\ref{s:conclusion} concludes with some remarks concerning the
renormalization process.

There have also been a number of attempts to calculate some of the required
renormalization factors numerically~\cite{Mandula:1997hb,Hashimoto96}.  The
renormalization factors computed from numerical simulations should agree with
the perturbative calculations as the weak coupling limit is approached.  This
is an important test of the numerical renormalization techniques, which has not
yet been attempted.  The renormalization of the current has never been computed
numerically.


\section{Continuum HQET}
\label{s:continuum}

Heavy Quark Effective Theory is a way of studying a single heavy quark in a
hadron when the mass of the quark is much larger than $\Lambda_{QCD}$.  See
Neubert~\cite{Neubert94a} for a nice review of HQET\@.  Mannel~{\it et
al.}~\cite{Mannel92} make rigorous Georgi's~\cite{Georgi90} intuition that the
heavy quarks at different velocities do not interact.  They do so by showing
that the QCD Green functions which involve two heavy quarks at different
velocities go to zero in the infinite mass limit.
So, there is a separate field for each heavy quark at each velocity.  In
HQET~\cite{Neubert94a}, the connection between the HQET fields and the quark
fields in QCD (Q) is:
\begin{equation}  \label{eq:h-defn}
\begin{array}{lcc}
           & h_v(x) = e^{iM v \cdot x} P_+ Q(x) &  \\
           & H_v(x) = e^{iM v \cdot x} P_- Q(x) &
\end{array}
\end{equation}
where $P_\pm = {}^{1}\!\!/\!_{2} ( 1 \pm v\!\!/ )$.  The new form of the QCD
Lagrangian has $h$ describing massless degrees of freedom and $H$ describing
fluctuations with twice the heavy quark mass.  Further, explicit Gaussian
integration of the $H$ fields produces the effective, non-local Lagrangian.
Upon integrating out the heavy degrees of freedom, the $H$ term is replaced by
a local term involving the light degrees of freedom, $h$, and the mass of the
heavy quark, $M$.  The Lagrangian is then expanded in the reciprocal of the
heavy quark mass; the zeroth order HQET Lagrangian is
\begin{equation} \label{eq:Leff}
{\cal L}_{\rm eff}
= \bar{h}_v i v \cdot D h_v,
\end{equation}
with the additional terms treated perturbatively as higher order in the
reciprocal of the heavy quark mass.  At zeroth order, i.e.\ in the infinite
mass limit, the theory is independent of the mass of the heavy quark, and the
Isgur-Wise function is universal (flavor-blind).

In HQET, the momentum of the heavy quark ($M_Q v$) is distinguished from the
momentum of the light quarks and gluons ($k$, the ``residual momentum''):
\[ M v = M_Q v + k. \]
The residual momentum is the difference between the momentum of the hadron ($M
v$) and the momentum of the heavy quark.  The velocity of the heavy quark
becomes a parameter of the theory and it is the residual momentum which becomes
conjugate to the position.  In the infinite mass limit, the momentum of the
hadron is due only to the heavy quark.

The matrix element in the continuum $\overline{MS}$ scheme is connected to the
matrix element calculated on the lattice by~\cite{Sachrajda:1988ak}
\begin{equation}
 \langle v  \mid V_{\mu} \mid v' \rangle^{\overline{MS}}  =
\frac{ Z_{\xi}^{\rm c} } { Z_{\xi}^{\rm l} }
 \langle v  \mid V_{\mu} \mid v' \rangle^{\rm latt}   =
       Z_{\xi}^{\rm cl}
 \langle v  \mid V_{\mu} \mid v' \rangle^{\rm latt}
\label{eq:keymatching}
\end{equation}
where $Z_{\xi}^{\rm c}$ is a continuum perturbative factor, $Z_{\xi}^{\rm l}$
is the lattice perturbative factor, and $Z_{\xi}^{\rm cl}$ is the ratio of the
two.
Falk~{\it et al.}~\cite{Falk90} calculated the continuum renormalization
factor:
\begin{equation}
\label{eq:cont renorm}
Z_{\xi}^{\rm c} = 1 +
\frac{g^2}{12\pi^2}
\left\{ 2 \left[ 1 - (v \cdot v') r(v \cdot v') \right]
       \ln (\mu / \lambda)^2
       +\delta_c   \right\},
\end{equation}
where
\begin{equation}  \label{eq:rfalk}
r(w) = \frac{\ln\!\left(w+\sqrt{w^2-1}\right)}{\sqrt{w^2-1}}
\end{equation}
and $\lambda$ is the gluon mass introduced as a infrared regulator. The
dependence on $\lambda$ must cancel in $Z_\xi^{\rm cl}$, the ratio of
$Z_\xi$'s, of Eq.~(\ref{eq:keymatching}).  In the $\overline{MS}$ scheme,
$\delta_c = 0 $~\cite{Neubert92}.  The calculation of the lattice
renormalization factor, $Z_{\xi}^{\rm l}$, is the subject of the next section.
$Z_{\xi}^{\rm cl}$ will be discussed further in Sec.~\ref{s:matching} when we
discuss the matching from the lattice to the continuum.


\section{Lattice HQET}
\label{s:lattice}

The Euclidean formulation of the lattice HQET action was introduced by Mandula
and Ogilvie~\cite{Mandula92}.
\begin{eqnarray}
S & = & \sum_x
        \left\{ v_0 \psi^\dagger\!\left(x\right)
                    \Delta_{t} \psi\!\left(x\right)
              - i \sum_j v_j \psi^\dagger\!\left(x\right)
                    \frac{\Delta_{j} + \Delta_{-j}}{2} \psi\!\left(x\right)
                \right\}.
\label{eq:mo;action-tad}
\end{eqnarray}
There is some freedom in the choice of which lattice derivatives are used in
Eq.~(\ref{eq:mo;action-tad}).
The tadpole improved finite differences are defined by
\begin{eqnarray}
\Delta_{\mu} \psi_{\vec{x}}
& = & \frac{U_{\vec{x},\vec{x}+\hat{\mu}}}{u_0} \psi_{\vec{x}+\hat{\mu}}
    - \psi_{\vec{x}}
      \label{eq:forward} \\
\Delta_{-\mu} \psi_{\vec{x}}
& = & \psi_{\vec{x}}
    - \frac{U_{\vec{x},\vec{x}-\hat{\mu}}^\dagger}{u_0}
\psi_{\vec{x}-\hat{\mu}}
      \label{eq:backward}
\end{eqnarray}
such that
\begin{center}
\begin{tabular}{cl}
$\psi^{\dagger}_{\vec{x}t} \Delta_{\mu} \psi_{\vec{x}t}$
& is a forward difference, \\
$\psi^{\dagger}_{\vec{x}t} \Delta_{-\mu} \psi_{\vec{x}t}$
& is a backward difference, \\
$\frac{1}{2}
 \left( \psi^{\dagger}_{\vec{x}t} \Delta_{\mu} \psi_{\vec{x}t} +
        \psi^{\dagger}_{\vec{x}t} \Delta_{-\mu} \psi_{\vec{x}t} \right)$
& is a centered difference,
\end{tabular}
\end{center}
and $u_0$ is the tadpole improvement factor~\cite{Lepage93}.  The tadpole
renormalization of the lattice HQET action is subtle because of the constraint
on the velocity; these subtleties are addressed in Appendix~\ref{a:tadpole}.

The centered difference approximates the continuum derivative to $O(a^2)$
(where $a$ is the lattice spacing); both the forward and backward difference
derivatives have $O(a)$ corrections to the continuum. Therefore, it seems that
the centered difference is the preferred type of derivative. This is true for
the spatial derivative; however, Mandula and
Ogilvie~\protect{\cite{Mandula:1997hb}} emphasize that for consistency an
asymmetric time difference must be employed, rather than a centered difference.
If a centered difference is employed, then the propagator vanishes on alternate
sites in the positive time direction and there is no continuum limit.  The
source of this problem is that the heavy quark fields are defined separately
from the heavy antiquark fields and are distinct for each velocity
[recall~Eq.~(\ref{eq:h-defn})]; thus, heavy quarks can only propagate in one
temporal direction.

The lattice HQET action originally proposed by Mandula and
Ogilvie~\cite{Mandula92} used a forward time derivative.  The backward time
derivative can be less convenient for use in simulations because a three
dimensional matrix must be inverted for each time step. The forward time
derivative only requires a matrix multiplication at each time step, and so is
computationally cheaper to simulate.

This choice of a forward time derivative has also been discussed by Davies and
Thacker~\cite{Davies92} in the context of NRQCD\@.  However, recent NRQCD
calculations follow the prescription of Lepage~{\it et
al.}~\cite{Lepage:1992tx} who use a backward time derivative but avoid having
to invert a large spatial matrix by splitting the spatial part of the action
over two adjacent time slices.  Their action, which can be ${\cal O}(a)$
improved, is symmetric with respect to time reversal, yet avoids the problems
of the centered difference.  Improved heavy-Wilson
actions~\cite{El-Khadra:1997mp} also go over to the backward derivative in the
static limit.  Similarly, better choices for the HQET lattice actions can be
made, and if we were to rerun the program with higher-order corrections, it
would indeed be advantageous to use the backwards time derivative as is done
for heavy-Wilson and modern NRQCD actions.  But for our present purposes, the
zeroth-order action suffices, and at this order the forward difference provides
a technical advantage in computation.

Since Aglietti's~\cite{Aglietti94} perturbative calculation used the
backward-difference time derivative, we do the perturbative calculations for
both types of time derivative.  We can check our results against Aglietti's,
against the results from the static theory~\cite{Eichten90c}, and also the
static limit of NRQCD~\cite{Davies92}.  Comparison in perturbation theory
between the forward and backward difference actions for the static case has led
to the introduction of the ``reduced wavefunction renormalization'' discussed
in Sec.~\ref{ss:reduced} and summarized in Appendix~\ref{a:reduced}.  (Please
see Appendix~\ref{a:notation} for a comparison of the notation between the
groups.)

We introduce the notation
\begin{equation}
\label{eq:fdbd}
\sigma = \left\{ \begin{array}{ll}
+1 & \mbox{forward difference} \\
-1 & \mbox{backward difference} \end{array} \right.
\end{equation}
in order to compare the forward versus backward difference actions.  Both the
forward and backward difference actions can be represented simultaneously by
replacing $\Delta_{t}$ by $\Delta_{\sigma t}$, where $\Delta_{\sigma t}$ is
either a forward time difference or a backward time difference, depending on
the choice of action.

Feynman rules can be derived from the action:
\begin{eqnarray}
\begin{array}{c} \mbox{quark} \\ \mbox{propagator} \end{array}
&& \left[ v_0 \sigma \left( \frac{1}{u_0} e^{i\sigma p_4} - 1 \right)
       + \sum_j \frac{v_j}{u_0} \sin(p_j) \right]^{-1}
\label{eq:feyn;prop} \\
\begin{array}{c} \mbox{gluon} \\ \mbox{propagator} \end{array}
&& \Delta(k)
= \left[ \sum_\mu 4 \sin^2\frac{k_\mu}{2} + \lambda^2 a^2 \right]^{-1}
\label{eq:feyn;gluon} \\
\mbox{vertex\ \ \ \ }
& \phantom{sp}
& \left[ \delta_{\mu,0} \left( i g (T^a)^{bc} \frac{v_0}{u_0}
                               e^{i\sigma(2p_4+k_4)/2} \right)
       + \sum_j \delta_{\mu,j} \left( g (T^a)^{bc} \frac{v_j}{u_0}
                                      \cos\frac{2p_j+k_j}{2} \right)
         \right]
  \label{eq:feyn;vert} \\
\begin{array}{c} \mbox{tadpole} \\ \mbox{vertex} \end{array} \ \
&& \left[ \delta_{\mu,0} \left(
\sigma\frac{g^2}{2}\frac{v_0}{u_0}(T^a)^{bd}(T^a)^{dc}
                               e^{i\sigma p_4} \right)
       + \sum_j \delta_{\mu,j}
\left(\frac{g^2}{2}\frac{v_j}{u_0}(T^a)^{bd}(T^a)^{dc}
                                      \sin{p_j} \right)
         \right]
  \label{eq:feyn;tad} \\
\begin{array}{c} \mbox{internal} \\ \mbox{integrations} \end{array}
&& \int_{-\pi/a}^{\pi/a} \frac{d^4k}{(2\pi)^4}
  \label{eq:feyn;int}
\end{eqnarray}
The $T^a$ are the color generators and $C_F=\frac{4}{3}$ is the Casimir
invariant.  $\lambda$ is a gluon mass, which is needed to regulate the infrared
divergences (as is done in the continuum) and which will be taken to zero at
the end of the calculation.

{}From the Feynman rules, it is straightforward to derive the usual self-energy
[$\Sigma(p)$], tadpole [$\Sigma^{\rm tad}(p)$], and vertex [$V(p,p')$]
corrections [the self-energy is $\Sigma(p) + \Sigma^{\rm tad}(p)$]:
\begin{eqnarray}
\Sigma(p)
& = & g^2 C_F \int \frac{d^4k}{(2\pi)^4} \frac{1}{\Delta(-k)}
      \frac{- \frac{v_0^2}{u_0^2} e^{i\sigma(2p_4+k_4)}
            + \sum_j \frac{v_j^2}{u_0^2} \cos^2\frac{2p_j+k_j}{2} }
           { \left[ v_0 \sigma \left( \frac{1}{u_0}
                                      e^{i\sigma (p_4+k_4)} - 1 \right)
                  + \sum_j \frac{v_j}{u_0} \sin(p_j+k_j) \right] }
      \label{eq:int;self4} \\
&& \nonumber \\
\Sigma^{\rm tad}(p)
& = &-\frac{g^2C_F}{2}
      \left(-\sigma \frac{v_0}{u_0} e^{i\sigma p_4}
           - \sum_j \frac{v_j}{u_0} \sin(p_j) \right)
      \int \frac{d^4k}{(2\pi)^4} \frac{1}{\Delta(k)}
      \label{eq:int;tad} \\
&& \nonumber \\
V(0,0)
& = & g^2 C_F \int \frac{d^4k}{(2\pi)^4} \frac{1}{ \Delta(-k) }
      \left\{- \frac{v_0 v_0'}{u_0^2}
              e^{i\sigma k_4/2} e^{i\sigma k_4/2}
            + \sum_j \frac{v_j v'_j}{u_0^2}
              \cos\frac{k_j}{2} \cos\frac{k_j}{2}
              \right\} /
      \nonumber \\ && /
      \left\{\left[ v_0 \sigma \left( \frac{e^{i\sigma k_4}}{u_0} - 1 \right)
                  + \sum_j \frac{v_j}{u_0} \sin(k_j) \right]
             \left[ v'_0 \sigma \left( \frac{e^{i\sigma k_4}}{u_0} - 1 \right)
                  + \sum_j \frac{v'_j}{u_0} \sin(k_j) \right]
             \right\}
      \label{eq:int;vert}
\end{eqnarray}
It is sufficient for our purposes to evaluate the vertex function with zero
external momentum.  The explicit $p$ dependence is kept in the self-energy
since the derivative will be considered.  The integral which appears in the
tadpole correction is standard and has the value $\frac{1}{(2\pi)^4} \int d^4k
/ \Delta(k) = 0.154933$.

The evaluation of the integrals is non-standard because of the problems caused
by the spectrum of the Euclidean HQET action not being bounded from below. We
follow the formalism developed by Aglietti~\cite{Aglietti94} and by Mandula and
Ogilvie~\cite{Mandula:1997hb}, in which we must first perform the $k_4$
integration analytically and do so by transforming to $z$-space ($z = e^{\pm
ik_4}$). A contour is chosen that enforces the forward propagation of the HQET
quarks~\cite{Mandula:1997hb} as described below. (The connection to Minkowski
space via a Wick contraction is discussed by Aglietti~\cite{Aglietti94}.)  The
resulting three dimensional integrals are then calculated numerically.  (All of
the numeric integrations were computed with the {\sc vegas}
routine~\cite{Lepage78}.)

The analytic $k_4$ integration of Eqs.~(\ref{eq:int;self4})
through~(\ref{eq:int;vert}) reduces the four dimensional integration to a three
dimensional integration.  It is, however, more convenient to do this as a
contour integration in $z$-space~\cite{Mandula:1997hb} after an
action-dependent change of variables
\begin{eqnarray}
z = e^{i\sigma k_4}. \label{eq:ktoz}
\end{eqnarray}
For this change of variables, the gluon propagator is written
\begin{equation}\label{eq:delta3}
\Delta\!\left(k_\mu\right)
 = 2 \sum_\mu ( 1 - \cos k_\mu ) + ( a \lambda )^2
\ =\ 2\ ( 1 - \cos k_4 ) + \Delta_3(\vec{k})
\end{equation}
which defines $\Delta_3(\vec k)$.

The $k_4$ contour (along the real axis) transforms into the unit circle in
complex $z$-space.  A subtlety arises when deciding which poles to enclose by
the contour.  The quark propagator pole appears as
\begin{equation}
z_Q = e^{-i\sigma p_4}
      \left( u_0 - \sigma \sum_j \tilde{v}_j  \sin(p_j+k_j) \right).
\end{equation}
The gluon propagator poles appear at
\begin{equation}   \label{eq:z-sigma}
z_{\pm} = 1 + \frac{\Delta_3}{2}
          \pm \frac{1}{2} \sqrt{\Delta_3^2 + 4 \Delta_3}
\end{equation}
where $\Delta_3$ is defined by Eq.~(\ref{eq:delta3}).  The contour separates
the gluon poles.  The contour should enclose the quark pole and one of the
gluon poles.  The subtlety is in choosing which gluon pole.  Because the
energy-momentum relation from the quark propagator, Eq.~(\ref{eq:e-mom}) of
Appendix~\ref{a:reduced}, can be negative, we split $k$-space (or $z$-space)
into a positive-energy region and a negative-energy region and enclose the
gluon pole which lies in the positive energy region of the space.
For negligible external momentum with a quark momentum $p+k$, the upper
$k_4$ half-plane is positive energy and, using Eq.~(\ref{eq:fdbd}) to
distinguish the actions, it is convenient to define $z$ via
Eq.~(\ref{eq:ktoz}) such that $z=e^{+ik_4}$ for the forward difference
action and $z=e^{-ik_4}$ for the backward difference action.
(For $p-k$, the lower $k_4$ half-plane is positive energy and it is convenient
to use $z = e^{-i \sigma k_4}$.)
With either of these choices, the backward difference action will have the
positive energy region outside of the $z$-space unit circle and the forward
difference action will have the positive energy region inside the $z$-space
unit circle.

The quark pole,
\begin{equation}
z_Q \sim
      \left( 1 - \sigma \frac{\vec{v}\cdot\vec{k}}{v_0} \right),
\end{equation}
with positive-energy [using Eq.~(\ref{eq:e-mom}), $z_Q \approx 1 - \sigma
\varepsilon$] is just inside (outside) of the unit circle for the forward
(backward) difference action.  Since $\sqrt{\Delta_3^2 + 4 \Delta_3} \geq
\Delta_3$, we find $z_+$ {\em outside} (and $z_-$ {\em inside}) of the unit
circle.  $z_\sigma$ (which is equal to $z_+$ for the forward difference and
$z_-$ for the backward difference) is therefore in the negative energy region.
Since the $k_+$ ($z_+$) gluon pole is always in the positive energy region for
the backward-difference action and always in the negative energy region for the
forward-difference action, we can write
\begin{equation}
\left.
\begin{array}{lcl}
\mbox{Backward Difference}
& \left\{ \begin{array}{c} z_+ \\ z_- \end{array} \right.
& \mbox{\begin{tabular}{r} positive energy pole \\ negative energy pole
\end{tabular}}
\\
\mbox{Forward Difference}
& \left\{ \begin{array}{c} z_+ \\ z_- \end{array} \right.
& \mbox{\begin{tabular}{r} negative energy pole \\ positive energy pole
\end{tabular}}
\end{array}
\right\}
\begin{array}{cr} z_\sigma & \mbox{negative energy pole} \\
                  z_{-\sigma} & \mbox{positive energy pole} \end{array}
\label{eq:gluon-pole}
\end{equation}
In both cases, it is the quark and the positive energy gluon poles which are
enclosed by the contour regardless of where the quark pole actually appears.
When the quark pole moves into the negative-energy region, it is necessary to
deform the contour to keep the quark pole enclosed.  (This is discussed for
$k$-space by Aglietti~\cite{Aglietti94} and for $z$-space by Mandula and
Ogilvie~\cite{Mandula:1997hb}.)  However, to simplify, one can equate this to
the negative of the contour integral which encloses only the negative energy
gluon pole.  The three-dimensional integrals resulting from the contour
integration have an action-dependent form due to the appearance of the negative
energy gluon pole ($z_\sigma$).  This pole is a function of $\vec k$.

In order to compute the renormalizations, the unrenormalized propagator is
compared to the renormalized propagator.  (We include the mass term in order to
calculate the mass renormalization.)  The renormalized propagator has the form
\begin{equation}
iH^{\rm r}(k)
 = \frac{Z}{\left[ i v^{\rm r}_0 k_4 + \sum_j v^{\rm r}_j k_j
                 + M^{\rm r} + O\!( k^2 ) \right]}.
\end{equation}
The renormalization factors are obtained by Taylor series expanding the
unrenormalized propagator:
\begin{eqnarray}
iH(k)
& = & \left[ v_0 \sigma \left( \frac{1}{u_0} e^{i\sigma k_4} - 1 \right)
           + \sum_j \frac{v_j}{u_0} \sin(k_j) + M_0 - \Sigma(k,v)
             \right]^{-1}.
\end{eqnarray}
We used
\begin{equation}
\Sigma(k,v) = \Sigma(0,v) + k_4 X_4 + \sum_j k_j X_j + O\!( k^2 )
\end{equation}
and
\begin{equation} \label{eq:trick}
\frac{1}{u_0} e^{i\sigma k_4}
 =  e^{i\sigma k_4 - \ln u_0}
 =  1 + i\sigma k_4 - \ln u_0 + O\!( k^2 )
\end{equation}
where
\begin{equation}
X_\mu = \left. \displaystyle \frac{\partial\, \Sigma\!(k)}{\partial
k_\mu\hfill} \right|_{k=0}.
\end{equation}
Equation~(\ref{eq:trick}) was used for the static
case~\cite{Bernard:1994zh,Bernard94a} to elucidate that the tadpole factor
$u_0$ results in mass renormalization rather than wavefunction renormalization.
Notice that $u_0$ has the perturbative expansion $[1-\frac{g^2 C_F}{(4\pi)^2}
\pi^2 + O\!( g^4 )]$, so $\ln u_0 \sim O\!( g^2 )$; the higher order terms are
neglected.  After a little algebra which involves the addition and subtraction
of some deducible terms, one can write the propagator in the form
\begin{eqnarray}
iH(k)
& = & \left\{ \left[ 1 - \delta Z \right]
              \left[ i \left( v_0 + \delta v_0 \right) k_4
                   + \sum_j \left( \frac{v_j}{u_0} +
\delta{}^{\underline{v_j}}_{u_0} \right) k_j
                   + \left( M_0 + \delta M \right)
                   + O\!( k^2 )
                     \right] \right\}^{-1},
\end{eqnarray}
which implies the expressions for the renormalizations:
\begin{eqnarray}
\delta M
\ = & M^{\rm r} - M_0
& = \ - \Sigma(0,v) - \sigma v_0 \ln u_0
      \label{eq:ren;mass} \\
\delta Z
\ = & Z - 1
& = \ - i v_0 X_4 - u_0 \sum_j v_j X_j
      \label{eq:ren;wave} \\
\delta{}^{\underline{v_i}}_{u_0}
\ = & v^{\rm r}_i -  \displaystyle \frac{v_i}{u_0}
& = \ - i v_0 \frac{v_i}{u_0} X_4 - (1 + v_i^2) X_i
      - v_i \sum_{j \ne i} v_j X_j
      \label{eq:ren;velj} \\
\delta v_0
\ = & v^{\rm r}_0 - v_0
& = \ - i (v_0^2 - 1) X_4 - v_0 u_0 \sum_j v_j X_j
    \label{eq:ren;vel0}
\end{eqnarray}
We make the following points regarding these expressions: First, in the HQET
formalism, the residual momentum is conjugate to the position, leaving the
velocity as a free parameter.  As discussed by Aglietti~\cite{Aglietti94}, the
velocity is renormalized on the lattice.  In the continuum, the four-vector
$X_{\mu}$ is proportional to $v_{\mu}$, the only available four-vector; this
implies that there is no velocity renormalization.  On the lattice, with
reduced rotational symmetry, this is not the case.  Secondly, if $u_0$ is set
to unity and the special case of $\vec{v} = v_z \hat z$ is taken, then these
reduce, for the backward-difference case, to Aglietti's
result~\cite{Aglietti94}.  Thirdly, $\delta{}^{\underline{v_j}}_{u_0}$ is a
notation to remind the reader that this quantity renormalizes $\frac{v_j}{u_0}$
rather than $v_j$ as can be seen in Eq.~(\ref{eq:ren;velj}).  For $u_0=1$, our
$\delta{}^{\underline{v_j}}_{u_0}$ corresponds to Aglietti's $\delta v_j$.
Further, the velocity renormalization can be written as follows:
\begin{eqnarray}
v^{r,\rm tad}_j
  =   v_j^{b,\rm tad}Z_{v_j}^{\rm tad}
& \ \ \ \ &
v^{r,\rm tad}_0
  =   v^{b,\rm tad}_0 Z_{v_0}^{\rm tad}
      \nonumber \\
Z_{v_j}^{\rm tad}
  =   \frac{1}{u_0}
       \left( 1
            + \frac{\delta^{\underline{v_j}}_{u_0}}
                   {{}^{\underline{v_j}}_{u_0}} \right)
& \ \ \ \ &
Z_{v_0}^{\rm tad}
  =    1 + \frac{\delta v_0}{v_0}
       \label{eq:vel-renormalization}
\end{eqnarray}
Finally, the $u_0$ that appears in these expressions is the perturbatively
expanded $u_0=1-\frac{g^2C_F}{16\pi^2} \pi^2$.  It is taken at lowest order
(unity) and the terms higher order in $g^2$ are ignored because $X_\mu \sim
O\!( g^2 )$.  The result is that the wavefunction renormalization and the first
term of the mass renormalization, $\Sigma(0,v)$, are the same to $O\!( g^2 )$
whether or not one uses tadpole improvement.  The velocity renormalization is
affected by tadpole improvement as
\begin{eqnarray}
Z_{v_j}^{\rm tad}
& = & \left[ 1
           + \frac{\delta{}^{\underline{v_j}}_{u_0}}
                  {{}^{\underline{v_j}}_{u_0}}
           - \frac{g^2 C_F}{16 \pi^2} (-\pi^2)
             \right]
      \nonumber \\
& = & \left[ 1
           + \frac{g^2 C_F}{16 \pi^2} \left( c(\tilde{v}) + \pi^2 \right)
             \right]
      \label{eq:Zv}
\end{eqnarray}
where the ($-\pi^2$) is from the perturbative expansion of $u_0$ and
$c(\tilde{v})$ is the same [to $O\!( g^2 )$] regardless of tadpole improvement.

Of the renormalization factors [Eqs.~(\ref{eq:ren;mass})
to~(\ref{eq:ren;vel0})], only the mass renormalization,
Eq.~(\ref{eq:ren;mass}), depends explicitly on the choice of forward or
backward time difference (the $\sigma$ parameter).  However, all the
renormalization factors implicitly depend on $\sigma$ via the $X_\mu$
functions.  The explicit dependence of the mass renormalization on $\sigma$ is
zero when tadpole improvement is not used; this is discussed further in
Appendix~\ref{a:reduced}, above Eq.~(\ref{eq:Zreduced}).


\subsection{Velocity Renormalization}
\label{ss:velocity}

Mandula and Ogilvie~\cite{Mandula:1997hb} renormalize $\tilde{v}_j \equiv
\frac{v_j}{v_0}$ rather than $v$.  We will not be using their notation, rather
we will be renormalizing $v$, and calculating $c(\tilde{v})$ defined by
Eqs.~(\ref{eq:ren;velj}), (\ref{eq:vel-renormalization}), and~(\ref{eq:Zv}):
\begin{equation} \label{eq:ag-deltav}
\frac{\delta{}^{\underline{v_j}}_{u_0}}{{}^{\underline{v_j}}_{u_0}}
 = \frac{g^2 C_F}{16\pi^2} c(\tilde{v}_j).
\end{equation}
This parallels the notation of Aglietti~\cite{Aglietti94}.  (See
Appendix~\ref{a:notation} for a comparison.)  Recall that this is the
perturbative renormalization to the tadpole-improved velocity.  Neither Mandula
and Ogilvie nor Aglietti use a tadpole-improved action.

The expression for $c(\tilde{v})$ is found from the self-energy Feynman
diagrams as expressed through Eq.~(\ref{eq:ren;velj}).  Continuing to use the
$\sigma = \pm 1$ to distinguish between the actions, we find
\begin{eqnarray}
c(\tilde{v})
& = & \frac{2 v_0^2}{\pi}
      \int \frac{d^3k}{\sqrt{\Delta_3^2+4\Delta_3}}
      \left( \frac{-2 \sigma z_\sigma\!\left(k\right)
                  + \frac{\tilde{v}_i}{u_0}
                    ( \frac{1}{v_0^2} + \tilde{v}_i^2 ) \sin\!\left(k_i\right)
                  + \sum_{j \ne i} \frac{\tilde{v}_j^3}{u_0}
\sin\!\left(k_j\right)}
                  { \sigma (\frac{1}{u_0} z_\sigma\!\left(k\right)-1)
                  + \sum_j \frac{\tilde{v}_j}{u_0} \sin\!\left(k_j\right)}
             \right. \nonumber \\ & & \left.
           + \frac{ \left[ z_\sigma\!\left(k\right)
                         - \sum_j \frac{\tilde{v}_j^2}{u_0^2}
                           \cos^2\!\left(\frac{k_j}{2}\right) \right]
                    \left[ z_\sigma\!\left(k\right)
                         - \left( \frac{1}{v_0^2} + \tilde{v}_i^2 \right)
                           \cos\!\left(k_i\right)
                         - \sum_{j\ne i} \tilde{v}_j^2 \cos\!\left(k_j\right)
\right] }
                  { \left[ \sigma ( \frac{1}{u_0} z_\sigma\!\left(k\right)-1)
                         + \sum_j \frac{\tilde{v}_j}{u_0}
\sin\!\left(k_j\right) \right]^2}
             \right)
      \label{eq:vel-renorm}
\end{eqnarray}
The $u_0$ are perturbatively expanded such that at this order in $g^2$, they
can be replaced with unity. (They are included as a reminder that in the next
order there will be an effect.)  Note that $z_\sigma(k)$ is the negative energy
gluon pole, defined by Eqs.~(\ref{eq:z-sigma}) and~(\ref{eq:gluon-pole}),
introduced from the residue of the contour.

Mandula and Ogilvie~\cite{Mandula:1997hb} perform an expansion in small
velocity and present the velocity renormalization as coefficients to powers of
the velocity.  (This is convenient in that whenever a calculation at a new
velocity is desired, the value for the velocity has precalculated coefficients
so that the calculation need not be done repeatedly.)  While this is
straightforward for the velocity renormalization, the divergences in the
wavefunction renormalization and the vertex correction make this technique more
complicated for these other calculations.  However, if we consider the
expansion for the velocity renormalization, then we get consistent results at
$O\!( \tilde{v}^6 )$ (notice that our format is slightly different because $c$
renormalizes $v$ rather than $\tilde{v}$)
\begin{eqnarray}
c(\tilde{v})
& = & c_{000}
    + c_{200} \tilde{v}_i^2
    + c_{020} \sum_{j \neq i} \tilde{v}_j^2
      \nonumber \\ & &
    + c_{400} \tilde{v}_i^4
    + c_{220} \tilde{v}_i^2 \sum_{j \neq i} \tilde{v}_j^2
    + c_{040} \sum_{j \neq i} \tilde{v}_j^4
    + c_{022} \sum_{k \neq j,i} \sum_{j \neq i} \tilde{v}_j^2 \tilde{v}_k^2
      \nonumber \\ & &
    + c_{600} \tilde{v}_i^6
    + c_{420} \tilde{v}_i^4 \sum_{j \neq i} \tilde{v}_j^2
    + c_{240} \tilde{v}_i^2 \sum_{j \neq i} \tilde{v}_j^4
    + c_{222} \tilde{v}_i^2 \sum_{k \neq j,i} \sum_{j \neq i} \tilde{v}_j^2
\tilde{v}_k^2
      \nonumber \\ & &
    + c_{060} \sum_{j \neq i} \tilde{v}_j^6
    + c_{042} \sum_{k \neq j,i} \sum_{j \neq i} \tilde{v}_j^4 \tilde{v}_k^2
    + \ldots
      \label{eq:mo-coefv-6}
\end{eqnarray}
The forward and backward difference results of this expansion are listed in
Table~\ref{t:vel-renorm-coefs}.  Mandula's and Ogilvie's results are reproduced
by the first two columns.
\begin{table}
\caption{\label{t:vel-renorm-coefs}
         The coefficients, $c_{mnl}$, used in the velocity
         renormalization when expanded in powers of the velocity to
         $O\!( \tilde{v}^6 )$ according to
         Eq.~(\protect{\ref{eq:mo-coefv-6}}).  $s$ is the order of the
         velocity term, found by summing the indices: $s = m+n+l$.  The
         first set is for the forward difference action; the second set
         is for the backward difference action.  If you consider the
         velocity in only one direction, then only the top row is
         relevant.}
\begin{center}
%
%
\begin{tabular}{ccccccccccc} \hline \hline
\multicolumn{2}{l}{$c_{mnl}$} & \multicolumn{4}{c}{Forward Difference} &&
\multicolumn{4}{c}{Backward Difference} \\
$n$ & $l$ & $s$=0 & $s$=2 & $s$=4 & $s$=6 && $s$=0 & $s$=2 & $s$=4 & $s$=6 \\
\hline
0 & 0 & $-28.07(3)$ & $-4.977(6)$ & $-1.093(3)$ & $-0.458(2)$ && $11.78(1)$ &
$0.33(2)$ & $-0.88(3)$ &  $-2.03(3)$ \\
2 & 0 &             & $-4.292(6)$ & $-2.100(6)$ & $-1.380(6)$ &&            &
$10.26(2)$ &  $9.49(6)$ &   $7.0(2)$ \\
4 & 0 &             &             & $-1.010(3)$ & $-1.346(6)$ &&            &
         &  $7.62(3)$ &  $28.1(2)$ \\
2 & 2 &             &             & $-1.005(6)$ & $-1.36(1)$  &&            &
         &  $9.53(6)$ &  $43.4(3)$ \\
6 & 0 &             &             &             & $-0.469(2)$ &&            &
         &            &  $39.98(6)$  \\
4 & 2 &             &             &             & $-4.54(2)$  &&            &
         &            & $109.6(6)$ \\
\hline\hline
\end{tabular}
\end{center}
\end{table}
Our results for the same special case (backward difference, $v_{x}=v_{y}=0$)
that Aglietti considers~\cite{Aglietti94} are listed in Table~\ref{t:delta-v}
and agree with Aglietti where they overlap.  The three columns of the forward
difference are: $c(\tilde{v})$ according to Eq.~(\ref{eq:vel-renorm}), its
expansion through sixth-order in small velocity according to
Eq.~(\ref{eq:mo-coefv-6}), and its expansion through second-order of the
velocity expansion using only the first three terms of
Eq.~(\ref{eq:mo-coefv-6}).  The latter confirms Mandula's and Ogilvie's result;
however, the sixth order result (using the coefficients of
Table~\ref{t:vel-renorm-coefs}) is in much better agreement with the exact
result (as one would expect).  Although Table~\ref{t:delta-v} only considers
motion along a single axis, our more general results indicate that for the
forward difference action it is sufficient to use the velocity expansion to
sixth order.
\begin{table}[tbp]
\caption[Velocity Renormalization Comparison]
        {\label{t:delta-v}
         This table lists the velocity renormalization for both forward
         difference (our choice) and backward difference actions for
         the special case $v_{x}=v_{y}=0$.  The last two columns solve
         the expanded equation through the superscripted order.  The
         $c(\tilde{v})$ entries are exact, that is, not expanded in the
         velocity.  Note that the $\tilde{v} \rightarrow 0.0$ limit is
         considered even though there is no need to calculate the
         renormalization coefficient when $\tilde{v}=0$. ($c$ has no
         interpretation in the static limit.)}
\begin{center}
\begin{tabular}{ccccc} \hline \hline
& (Backward) & \multicolumn{3}{c}{(Forward)} \\
& $c(\tilde{v})$ & $c(\tilde{v})$ & $c^{(6)}$ & $c^{(2)}$ \\ \hline
$c(\tilde{v} \rightarrow 0.0)$
            & 11.779(4) & -28.06(1) & -28.06(1) &   ---     \\
$c(\tilde{v}=0.1)$ & 11.899(5) & -28.40(1) & -28.40(1) & -28.38(1) \\
$c(\tilde{v}=0.2)$ & 12.275(5) & -29.44(1) & -29.44(1) & -29.42(1) \\
$c(\tilde{v}=0.3)$ & 12.966(5) & -31.35(1) & -31.35(1) & -31.29(1) \\
$c(\tilde{v}=0.4)$ & 14.036(7) & -34.39(1) & -34.39(1) & -34.28(1) \\
$c(\tilde{v}=0.5)$ & 15.67(1)\ & -39.17(1) & -39.17(1) & -38.95(2) \\
$c(\tilde{v}=0.6)$ & 18.05(1)\ & -46.90(1) & -46.90(1) & -46.44(2) \\
$c(\tilde{v}=0.7)$ & 20.82(3)\ & -60.44(2) & -60.45(2) & -59.47(3) \\
\hline \hline
\end{tabular}
\end{center}
\end{table}

For the more general case of all the spatial velocities not equal to zero, we
present the results for the forward difference action at small velocities in
Table~\ref{t:my-c2}.
\begin{table}[p]
\caption[Velocity Renormalization for General Velocities]
         {\label{t:my-c2}
         The velocity renormalization, $c_z(\tilde{v})$, for the
         forward difference action for several general (small)
         velocities.  The uncertainty is at most 2 in the last digit.}
%
%
\begin{center}
\begin{tabular}{cccccc} \hline \hline
\\
$\tilde{v}_z=0.00$ & ${}_{\tilde{v}_x}\backslash^{\tilde{v}_y}$ & 0.00 & 0.05 &
0.10 & 0.25  \\
\cline{2-6}
 & 0.00 & $-28.06$ & $-28.15$ & $-28.40$ & $-30.14$ \\
 & 0.05 & $-28.16$ & $-28.23$ & $-28.47$ & $-30.19$ \\
 & 0.10 & $-28.39$ & $-28.47$ & $-28.72$ & $-30.46$ \\
 & 0.25 & $-30.12$ & $-30.20$ & $-30.45$ & $-32.17$ \\
\\
$\tilde{v}_z=0.05$ & ${}_{\tilde{v}_x}\backslash^{\tilde{v}_y}$ & 0.00 & 0.05 &
0.10 & 0.25  \\
\cline{2-6}
 & 0.00 & $-28.16$ & $-28.23$ & $-28.48$ & $-30.19$ \\
 & 0.05 & $-28.26$ & $-28.34$ & $-28.55$ & $-30.27$ \\
 & 0.10 & $-28.49$ & $-28.56$ & $-28.82$ & $-30.53$ \\
 & 0.25 & $-30.20$ & $-30.30$ & $-30.54$ & $-32.26$ \\
\\
$\tilde{v}_z=0.10$ & ${}_{\tilde{v}_x}\backslash^{\tilde{v}_y}$ & 0.00 & 0.05 &
0.10 & 0.25  \\
\cline{2-6}
 & 0.00 & $-28.40$ & $-28.48$ & $-28.72$ & $-30.45$ \\
 & 0.05 & $-28.49$ & $-28.55$ & $-28.81$ & $-30.53$ \\
 & 0.10 & $-28.72$ & $-28.80$ & $-29.06$ & $-30.78$ \\
 & 0.25 & $-30.45$ & $-30.52$ & $-30.79$ & $-32.50$ \\
\\
$\tilde{v}_z=0.25$ & ${}_{\tilde{v}_x}\backslash^{\tilde{v}_y}$ & 0.00 & 0.05 &
0.10 & 0.25  \\
\cline{2-6}
 & 0.00 & $-30.15$ & $-30.21$ & $-30.46$ & $-32.20$ \\
 & 0.05 & $-30.22$ & $-30.31$ & $-30.56$ & $-32.27$ \\
 & 0.10 & $-30.48$ & $-30.54$ & $-30.80$ & $-32.54$ \\
 & 0.25 & $-32.21$ & $-32.27$ & $-32.51$ & $-34.26$ \\
\\ \hline\hline
\end{tabular}
%
%
\end{center}
\end{table}
This is the factor, $c_z(\tilde{v})$, which renormalizes the $\hat{z}$
component of the velocity according to Eq.~(\ref{eq:Zv}).  The renormalizations
for the $v_x$ and $v_y$ components can be deduced from the table by symmetry.
Notice that the $v_z$ renormalization is affected by each component of
$\vec{v}$, not merely by $v_z$.  The numerical size of the perturbative factors
in Tables~\ref{t:delta-v} and~\ref{t:my-c2} are both large. Tadpole improving
the perturbative factors, by adding $+\pi^2$ to them as in Eq.~(\ref{eq:Zv}),
does not substantially reduce the size of the perturbative contribution.

To give an idea about the magnitude of the velocity renormalization, we
consider $\beta = 6.0$ with \mbox{$|\vec v| = 0.5$,} and use the bare lattice
coupling. The non-tadpole improved multiplicative factor $Z_{v_j} = 0.67$; the
corresponding tadpole improved number is $Z_{v_j}^{tad} = 0.75$. If the boosted
coupling, $g^2/u_0^4$~\cite{Lepage93}, is used then $Z_{v_j}^{tad} = 0.59$. As
the slope of the Isgur-Wise function essentially depends quadratically on the
velocity renormalization, this makes perturbation theory unreliable to analyze
the simulation data and thus numerical renormalization techniques must be
used~\cite{Christensen97b,Christensen98,Mandula:1997hb,Hashimoto96}.


\subsubsection{Aside: Slow HQET}
\label{sss:sHQET}

In Aglietti's~\cite{Aglietti94} initial calculations, the velocity
renormalization was presented as a function of the velocity. However, Mandula
and Ogilvie~\cite{Mandula:1997hb} expanded the velocity renormalization in a
power series in the velocity, which allowed them to compare their perturbative
results with the numbers from their numerical renormalization technique.  The
expansion of the renormalization factor in velocity seems to be similar to
Aglietti's~\cite{Aglietti:1993vz} idea of slow HQET, where the $v \cdot D$ term
is a perturbation on the static theory.  Slow HQET was studied in perturbation
theory by Aglietti and Gim\'enez~\cite{Aglietti95}, where they demonstrated
that slow HQET agreed with HQET in the infrared and ultraviolet limits. It
would be interesting to understand the connection between slow HQET and the
HQET formalism of Mandula and Ogilvie.

We have found expressions for the velocity renormalization in terms of the
coefficients for the backward difference action
(Table~\ref{t:vel-renorm-coefs}) and note that the $c_{042}$ coefficient of the
backward difference is rather large, at 109.6(6), -- much larger than the
equivalent coefficient for the forward difference action.  This could indicate
a problem with the expansion for the backward difference; the forward
difference coefficients, which we checked through $O\!( \tilde{v}^6 )$, are all
reasonably close to unity.  Aglietti and Gim\'enez do not calculate the sixth
order coefficient for the velocity renormalization (although they take the
other renormalizations to this order); however, the renormalization factors are
only required to quadratic order in the velocity in order to compute the slope
of the Isgur-Wise function from simulations of slow HQET.  (The slow HQET
formalism is used to directly calculate the derivatives of the Isgur-Wise
function, using the ``moments'' technique~\cite{Aglietti:1993vz}.  Aglietti and
Gim\'enez~\cite{Aglietti95} found that the expressions for the higher order
derivatives of the Isgur-Wise function, beyond the slope, contained operators
that diverged with an inverse power of the lattice spacing and that must be
subtracted off in the simulation.)


\subsection{Mass Renormalization}
\label{ss:mass}

As with the velocity renormalization, we define $x(\tilde{v})$ as the mass
renormalization without the $g^2$-prefactor.  For comparison,
Aglietti~\cite{Aglietti94} also does this; however, we prefer (for comparison
to the static limit of the forward difference NRQCD theory) to have our
$x(\tilde{v})$ proportional to $+\Sigma(0,\tilde{v})$.  So, our $x$ is the
negative of Aglietti's.  (See Appendix~\ref{a:notation} for a comparison
between groups.)  We also include the effect of tadpole improvement.
\begin{equation} \label{eq:x}
\delta M
= -\Sigma(0,\tilde{v}) - \sigma v_0 \ln u_0
= -\frac{g^2 C_F}{16\pi^2}
   \frac{x\!(\tilde{v})}{a}
  -\sigma v_0 \ln u_0
\end{equation}
Recall that $\ln u_0$ is $O\!( g^2 )$.  From Eq.~(\ref{eq:ren;mass}),
\begin{eqnarray}
x(\tilde{v})
& = & \frac{2v_0}{\pi}
      \int \frac{d^3k}{\sqrt{\Delta_3^2+4\Delta_3}}
      \frac{-z_\sigma\!\left(k\right)
           +
\sum_j\frac{\tilde{v}_j^2}{u_0^2}\cos^2\!\left(\frac{k_j}{2}\right)}
           { \left[ \sigma \left( \frac{1}{u_0} z_\sigma\!\left(k\right)
                                - 1 \right)
                  + \sum_j \frac{\tilde{v}_j}{u_0} \sin\!\left(k_j\right)
\right] }
    + \sigma 8\pi^2 v_0 (0.154933)
\end{eqnarray}
where the $8\pi^2 (0.154933)$ is from the tadpole contribution ($\Sigma^{\rm
tad}$) which is partially canceled by the second term in Eq.~(\ref{eq:x}) as it
should be.  The $u_0$ are again perturbatively expanded such that, at this
order, they can be replaced with unity.  They are included here as a reminder
that in the next order they will have an effect.
The values of this integral are listed in Table~\ref{t:my-ag} in the next
subsection.  As they are relevant to the reduced wavefunction, we will discuss
these there.


\subsection{Wavefunction Renormalization}
\label{ss:wavefunction}

The results of the wavefunction and reduced wavefunction renormalization can be
compared not only to Aglietti~\cite{Aglietti94} (backward difference, HQET) and
Mandula and Ogilvie~\cite{Mandula:1997hb} (forward difference, HQET), but also
to Eichten and Hill~\cite{Eichten90c,Eichten90b} (backward difference, static
theory) and the static limit of Davies and Thacker~\cite{Davies92} (forward
difference, NRQCD).  (Appendix~\ref{a:notation} compares the notations between
groups.)

Recall that the wavefunction renormalization can be found as
Eq.~(\ref{eq:ren;wave}).  During this calculation, as with the velocity
renormalization, the $k_0$ integration is done analytically with the same
comments as were made earlier.  This introduces the residue from the negative
energy gluon pole, $z_\sigma(k)$.  Again using the $\sigma = \pm 1$ to
distinguish between the actions, the result for $\delta Z$ is
\begin{eqnarray}
\delta Z
& = & \frac{g^2 C_F}{16 \pi^2} \frac{2 v_0^2}{\pi}
      \int \frac{d^3k}{\sqrt{\Delta_3^2+4\Delta_3}}
      \left\{ \frac{-2\sigma z_\sigma\!\left(k\right)
                  + \sum_j \frac{\tilde{v}_j^3}{u_0} \sin\!\left(k_j\right)}
                  { \left[ \sigma (\frac{1}{u_0} z_\sigma\!\left(k\right)-1)
                         + \sum_j \frac{\tilde{v}_j}{u_0}
\sin\!\left(k_j\right)
                           \right] }
      \right. \nonumber \\ & & \left. \phantom{\int}
    + \frac{ \left[ z_\sigma\!\left(k\right)
                  - \sum_j \frac{\tilde{v}_j^2}{u_0^2}
                    \cos^2\!\left(\frac{k_j}{2}\right) \right]
             \left[ z_\sigma\!\left(k\right)
                  - \sum_j \tilde{v}_j^2 \cos\!\left(k_j\right) \right] }
           { \left[ \sigma (\frac{1}{u_0} z_\sigma\!\left(k\right) - 1)
                  + \sum_j \frac{\tilde{v}_j}{u_0}
                           \sin\!\left(k_j\right) \right]^2 }
      \right\} \nonumber \\ & &
    + \frac{g^2 C_F}{16 \pi^2} 8\pi^2 (0.154933) \nonumber
\end{eqnarray}
where again the $8\pi^2 (0.154933)$ is from the tadpole contribution and the
$u_0$ are again perturbatively expanded such that at this order, they can be
replaced with unity.  This three-dimensional integral has a logarithmic
divergence.  The way with which this is typically dealt is to add and subtract
an integral with the same logarithmic divergence which is solvable
analytically.  We call this integral $\delta Z^c$ and use the small $k$ limit
because we are interested in the infrared (low energy) divergence.  The
difference $\delta Z - \delta Z^c$ is finite and calculated numerically.
$\delta Z^c$ (found analytically) will have a finite piece, which is added back
to the numerical calculation, as well as a divergent piece.  The divergent
piece contributes to the coefficient of the $\ln(\lambda^2a^2)$ term in the
renormalization of the lattice Isgur-Wise function.

Although the ``continuum-like'' limit of $\delta Z$ is actually
\begin{eqnarray}
&   & \frac{g^2 C_F}{16 \pi^2} \frac{2 v_0^2}{\pi}
      \int \frac{d^3k}{2 \sqrt{k^2+\lambda^2a^2}}
      \left\{ \frac{-2\sigma
                  + \sum_j \frac{\tilde{v}_j^3}{u_0} k_j }
                  { \left[ (\sqrt{k^2+\lambda^2a^2})
                         + \sum_j \frac{\tilde{v}_j}{u_0} k_j  \right] }
      \right. \nonumber \\ & & \left. \phantom{\int}
    + \frac{ \left[ 1
                  - \sum_j \frac{\tilde{v}_j^2}{u_0^2} \right]
             \left[ 1
                  - \sum_j \tilde{v}_j^2 \right] }
           { \left[  (\sqrt{k^2+\lambda^2a^2})
                  + \sum_j \frac{\tilde{v}_j}{u_0} k_j \right]^2 }
      \right\} \nonumber \\ & &
    + \frac{g^2 C_F}{16 \pi^2} 8\pi^2 (0.154933), \nonumber
\end{eqnarray}
the first and third terms are finite.  Since we are interested in the infrared
divergent piece, we will define $\delta Z^c$ as the second term.  By taking
advantage of $u_0 |_{\rm pert} = 1 + O\!( g^2 )$ as well as by using the
velocity normalization $\left[ \left( 1 - \sum_j \tilde{v}_j^2 \right) =
\frac{1}{v_0^2}\right]$ we get
\begin{eqnarray}
\delta Z^c
& = & \frac{g^2 C_F}{16 \pi^2} \frac{1}{\pi v_0^2}
      \int_0^R \frac{d^3k}{\sqrt{k^2+\lambda^2a^2}}
      \frac{ 1 }
           { \left[ \sqrt{k^2+\lambda^2a^2}
                  + \sum_j \tilde{v}_j k_j \right]^2 }.
      \nonumber
\end{eqnarray}
The upper limit $R$ is arbitrary because this term is added and subtracted.
Interestingly, this is the same integral for both actions.  The result of this
integral is (with $\tilde{v} = \sqrt{\sum_j \tilde{v}_j^2}$)
\begin{eqnarray}
&&
   \frac{g^2 C_F}{16\pi^2} \frac{1}{v_0^2}
   \left[ \frac{2}{1-\tilde{v}^2}
          \ln\!\left(\frac{\sqrt{R^2+\lambda^2a^2} + R}
                          {\sqrt{R^2+\lambda^2a^2} - R}\right)
        - \frac{2}{\tilde{v}(1-\tilde{v}^2)}
          \ln\!\left(\frac{\sqrt{R^2+\lambda^2a^2} + \tilde{v} R}
                          {\sqrt{R^2+\lambda^2a^2} - \tilde{v} R}\right)
          \right]
\nonumber \\
&& \ \begin{array}{c}\\^{\lambda \ll R^{\phantom{2}}}\end{array}
                        \!\!\!\!\!\!\!\!\!\!\!\!\!\!\!\!\!\!\!
                        -\!\!\!-\!\!\!-\!\!\!\longrightarrow
   \frac{g^2 C_F}{16\pi^2}
   \left[ 2
          \ln\!\left(\frac{4R^2}{\lambda^2a^2}\right)
        - \frac{2}{\tilde{v}}
          \ln\!\left(\frac{1 + \tilde{v}}{1 - \tilde{v}}\right)
          \right].
\label{eq:Zc}
\end{eqnarray}
The divergent piece is $-2\ln(\lambda^2a^2)$.  This is the wavefunction
contribution to the divergence in the renormalization of the Isgur-Wise
function.  We can, for convenience, set $R = {}^{1}\!\!/\!_{2}$.

As with the mass renormalization, the results for the wavefunction
renormalization are listed in Table~\ref{t:my-ag}, but discussed further in the
next subsection.  Note that, as with $x(\tilde{v})$ for the mass
renormalization, the wavefunction renormalization is referred to by
$e(\tilde{v})$ and defined by
\begin{equation}  \label{eq:e}
\delta Z = \frac{g^2 C_F}{16\pi^2}
           \left[ -2 \ln\!\left(\lambda^2a^2\right) + e(\tilde{v}) \right].
\end{equation}
%


\subsection{Renormalization of the Reduced Wavefunction}
\label{ss:reduced}

The perturbative factors for various heavy quark effective field theories
depend subtly~\cite{Aglietti94,Davies92,Eichten90c,Maiani92} on whether the
forward or backward time derivative is used in the action.  It is expected that
in the Euclidean formulation, the propagator as a function of time and the
residual 3-momentum (i.e.\ Fourier transforming $k_4$ into $t$) will have the
dependence $e^{-\varepsilon t} = e^{-mt}$.  However, it turns out
(App.~\ref{a:reduced}) to have the dependence $Ae^{-m(t-\sigma)}$ where
$\sigma=\pm1$ distinguishes the actions.  Eichten and Hill~\cite{Eichten90c}
noticed this relation and found that if one fits, instead, to $A'e^{-mt}$
(where $A' = A e^{m\sigma}$) this changes the wavefunction renormalization by
subtracting (or adding) the mass renormalization.  It also ``reduces'' the
wavefunction renormalization to a common answer for both the forward and
backward difference actions.  Since it is convenient to fit to $e^{-mt}$ and
the reduced value is the same for both actions, this is a popular choice.
Unfortunately, in lattice HQET away from the static limit the reduced values
(for the forward and the backward-difference cases) are not equal, as we will
show.

Equation~(\ref{eq:x}) defines $x(\tilde{v})$ in terms of the mass
renormalization.  Equation~(\ref{eq:e}) defines $e(\tilde{v})$ in terms of the
wavefunction renormalization.  Appendix~\ref{a:reduced} derives
Eq.~(\ref{eq:eprime}),
\begin{equation}
e'(\tilde{v}) \equiv e(\tilde{v}) - \sigma x(\tilde{v})/v^{\rm ren}_0,
\nonumber
\end{equation}
which is the relation for the reduced value of the wavefunction
renormalization.  The tadpole term is in $x(\tilde{v})$ [it gets canceled in
the mass renormalization of Eq.~(\ref{eq:x})] and, as noticed for the static
case in~\cite{Bernard:1994zh,Bernard94a}, the wavefunction and reduced
wavefunction renormalizations remain unaffected by tadpole improvement.
Table~\ref{t:my-ag} lists our values for these functions for a large velocity
range.  Notice that in the static limit, the reduced value for the two actions
is the same.  This is the expected result.  Also notice that our forward
difference value for $e(0.0)$ agrees with Davies and Thacker~\cite{Davies92}
(in their notation $C = Z + aA = -15.4$).  Our backward difference table agrees
with Aglietti~\cite{Aglietti94}; and the static limit of the backward
difference action, $e(0.0)=24.44$, is also in agreement with Eichten and
Hill~\cite{Eichten90c}.
\begin{table}[p]
\caption[Mass and Wavefunction Renormalizations]
        {\label{t:my-ag}
         Mass [$x(\tilde{v})$] and wavefunction [$e(\tilde{v})$]
         renormalization functions for the backward-difference (bd) and
         forward-difference (fd) actions.  The bd numbers reproduce
         Aglietti's table~\protect{\cite{Aglietti94}}.  The bd and fd
         numbers for $x(\tilde{v})$ and $e'(\tilde{v})$ should agree
         {\sl only} in the static limit ($\tilde{v}$=0).
         $e'(\tilde{v}) = e(\tilde{v}) - \sigma x(\tilde{v})/v_0$ are
         the reduced wavefunction. Notice that we define $x(\tilde{v})$
         as the negative of that of Aglietti.  In addition, $v_x = v_y
         = 0$.}
\begin{center}
\begin{tabular}{ccccccc} \hline \hline
& \multicolumn{3}{c}{Backward Difference} & \multicolumn{3}{c}{Forward
Difference} \\
$\tilde{v}$ & $x(\tilde{v})$ & $e(\tilde{v})$ & $e'(\tilde{v})$ &
$x(\tilde{v})$ & $e(\tilde{v})$ & $e'(\tilde{v})$ \\ \hline
0.0 & -19.92(3) & 24.43(4) & \ 4.53(1) & -19.93(1) & -15.40(1) &\ 4.530(4) \\
0.1 & -19.87(3) & 24.64(4) & \ 4.875(4)& -19.99(1) & -15.75(1) &\ 4.141(2) \\
0.2 & -19.69(3) & 25.24(4) & \ 5.97(1) & -20.17(1) & -16.82(1) &\ 2.935(2) \\
0.3 & -19.34(3) & 26.36(4) & \ 7.91(1) & -20.47(1) & -18.78(1) &\ 0.759(4) \\
0.4 & -18.75(3) & 28.14(4) &  10.96(1) & -20.97(1) & -21.91(1) & -2.694(6) \\
0.5 & -17.72(3) & 30.94(5) &  15.60(2) & -21.72(2) & -26.83(2) & -8.015(8) \\
0.6 & -15.79(3) & 35.44(5) &  22.82(4) & -22.89(1) & -34.74(2) & -16.44(2) \\
0.7 & -11.15(3) & 44.2(1)\ &  36.27(10)\ & -24.79(2) & -48.56(4) & -30.85(2)
\\ \hline \hline
\end{tabular}
%
%
\end{center}
\end{table}
While it is still convenient to use the reduced result and fit to $e^{-mt}$ in
the static limit, the forward and backward difference will have different
reduced wavefunction renormalizations away from the static limit.


\subsection{Vertex Correction}
\label{ss:vertex}

The vertex correction also has differences between the actions and a divergence
which must be subtracted as was done for the wavefunction renormalization.
However, this has the further complication that it depends on the velocities of
both the incoming and the outgoing quarks.  So whereas the wavefunction
renormalization is a function of $\tilde{v}$, the vertex correction, $\delta
V\!\left(\tilde{v},\tilde{v}'\right)$, is a function of the initial and final
velocities.

After analytically doing the contour integration over the $k_4$ variable and
dealing with the poles as discussed previously, we find
\begin{eqnarray}
\delta V\!\left(\tilde{v},\tilde{v}'\right)\!
& = & \frac{g^2 C_F}{16\pi^2} \frac{2}{\pi}
      \int \frac{d^3k}{\sqrt{\Delta_3^2+4\Delta_3}}
      \left\{-\frac{v_0 v_0'}{u_0} z_\sigma\!\left(k\right)
            + \sum_j \frac{v_j v_j'}{u_0^2}
              \cos^2\!\left({}^{k}\!\!/\!_{2}\right) \right\}
      \nonumber \\ & & /
      \left\{ \left[ v_0 \sigma \left( \frac{1}{u_0}
                                  z_\sigma\!\left(k\right) - 1 \right)
                   + \sum_j \frac{v_j}{u_0} \sin\!\left(k_j\right) \right]
              \left[ v_0' \sigma \left( \frac{1}{u_0}
                                   z_\sigma\!\left(k\right) - 1 \right)
                   + \sum_j \frac{v_j'}{u_0} \sin\!\left(k_j\right) \right]
\right\}.
       \label{eq:deltaV}
\end{eqnarray}
The $u_0$ are once again perturbatively expanded such that at this order, they
can be replaced with unity. (They are included here as a reminder that in the
next order they will have an effect.) For the rest of this subsection, we will
explicitly set $u_0$=1.

As claimed by Aglietti~\cite{Aglietti94}, this must have the form
\begin{equation}  \label{eq:d-defn}
\frac{g^2 C_F}{16\pi^2} \frac{2}{\pi}
\left[ 2 \, (v \cdot v') \, r\!\left(v \cdot v'\right) \,
\ln\!\left(\lambda^2a^2\right)
     + d(\tilde{v},\tilde{v}') \right].
\end{equation}
The lattice coefficient must have this form if it is to cancel the continuum
divergence $\ln\!\left({}^{\mu}\!\!/\!_{\lambda}\right)$ (which was computed by
Falk~{\it et al.}~\cite{Falk90}).  Of primary interest is that it be a function
of $v \cdot v'$, the only non-trivial invariant constructible from the heavy
quark velocities $v$ and $v'$.  We find that numerically the lattice divergence
coefficient agrees with the continuum divergence coefficient --- these are
listed in Table~\ref{t:lattice-falk}.
\begin{table}[p]
\caption[Coefficient of the Lattice Vertex Divergence]
         {\label{t:lattice-falk}
         The coefficient of the lattice divergent $\ln\!\left(\lambda
         a\right)$ piece must and does reproduce the continuum
         divergent coefficient [$4 \, (v \cdot v') \, r(v \cdot v')$] in
         order to correctly cancel the $\ln\lambda$.  Errors are at
         most three in the last digit shown.  In addition, $v_x = v_y =
         v'_x = v'_y = 0$.}
\begin{center}
\begin{tabular}{ccccccccc} \hline \hline
${}_{\tilde{v}^\prime}\backslash^{\tilde{v}}$
& 0.00 & 0.10 & 0.20 & 0.30 & 0.40 & 0.50 & 0.60 & 0.70 \\ \hline \hline
0.00 & $4.001$ & $4.014$ & $4.050$ & $4.127$
     & $4.237$ & $4.396$ & $4.618$ & $4.953$ \\
0.10 & $4.012$ & $4.004$ & $4.015$ & $4.059$
     & $4.139$ & $4.269$ & $4.462$ & $4.753$ \\
0.20 & $4.053$ & $4.011$ & $4.002$ & $4.018$
     & $4.064$ & $4.160$ & $4.319$ & $4.573$ \\
0.30 & $4.126$ & $4.058$ & $4.014$ & $4.000$
     & $4.016$ & $4.079$ & $4.194$ & $4.412$ \\
0.40 & $4.238$ & $4.137$ & $4.066$ & $4.016$
     & $4.000$ & $4.021$ & $4.095$ & $4.257$ \\
0.50 & $4.398$ & $4.267$ & $4.159$ & $4.077$
     & $4.022$ & $4.000$ & $4.026$ & $4.133$ \\
0.60 & $4.622$ & $4.460$ & $4.316$ & $4.197$
     & $4.095$ & $4.028$ & $3.998$ & $4.040$ \\
0.70 & $4.956$ & $4.753$ & $4.571$ & $4.412$
     & $4.257$ & $4.135$ & $4.041$ & $4.000$ \\
\hline\hline
\end{tabular}
%
%
\end{center}
\end{table}

Aglietti only gives results for the $\vec{v}=v_z\hat{z}$ with $\tilde{v}'=0$
case.  We have found a problem with an equation which he uses and have
introduced a better expression for finding the divergence.  We have also
extended the calculation to the forward difference action and to nonzero
$\tilde{v}'$.


\subsubsection{Vertex Correction with $\tilde{v}' \rightarrow 0$}

In the $\tilde{v}' \rightarrow 0$ limit, we can reproduce Aglietti's results;
the numerical values are listed in the next subsection.  Originally, we could
not satisfactorily reproduce Aglietti's numbers, especially for the
$\tilde{v}=0.0$ result.  As we investigated this, we found a problem with the
$\lambda \rightarrow 0$ limit, specifically there was a subtlety with the
interchange of limits ($\lambda \rightarrow 0$ versus $\tilde{v} \rightarrow
0$).  We believe that Aglietti's choice of integral subtraction can be
improved.  This subsection discusses this subtlety.
Tables~\ref{t:my-d-backward}, \ref{t:my-d-forward}, and~\ref{t:my-ag2} were
produced with our choice.

Since the $\delta V$ integral is divergent, a technique similar to that used
for $\delta Z$ can be used; however, it needs to be modified because the
``continuum-like'' limit of this integral, $\delta V^c$, is not analytically
manageable.  However, a second integral, $\delta V^{cc}$, can be taken such
that $\delta V - \delta V^c$ and $\delta V^c - \delta V^{cc}$ are each finite.
These numeric integrals are then done separately and added together along with
the finite piece of $\delta V^{cc}$.

Aglietti refers to $\delta V^c$ as $I$, and our $\delta V^{cc}$ is analogous to
his $L$.  Aglietti uses the notation $\delta(\tilde{v}) = L-I$; we will use the
analogous definition $\delta'(\tilde{v}) = \delta V^{cc} - \delta V^{c}$.  To
be explicit, in the small-$k$ limit Eq.~(\ref{eq:deltaV}) reduces to
\begin{eqnarray}
\delta V^c\!\left(\tilde{v},\tilde{v}'\right)
& = & \frac{g^2 C_F}{16\pi^2} \frac{2}{\pi}
      \int \frac{d^3k}{2\sqrt{k^2+\lambda^2a^2}}
           \frac{-1 + \sum_j \tilde{v}_j \tilde{v}_j' }
                { \left[\sqrt{k^2+\lambda^2a^2}+\sum_j\tilde{v}_j k_j \right]
                  \left[\sqrt{k^2+\lambda^2a^2}+\sum_j\tilde{v}_j'k_j \right] }
 \label{eq:deltaVc} \\
\delta V^c\!\left(\tilde{v},0\right)
& = & \frac{g^2 C_F}{16\pi^2} \frac{2}{\pi}
      \int \frac{d^3k}{2 \left(k^2+\lambda^2a^2\right)}
           \frac{-1}
                {\left[ \sqrt{k^2+\lambda^2a^2}+\sum_j\tilde{v}_jk_j\right]}
\label{eq:deltaVc0}.
\end{eqnarray}
Aglietti approximates $\delta V^c(\tilde{v},\tilde{v}'=0)$ with
\begin{eqnarray}
L = \frac{g^2 C_F}{16\pi^2} \frac{2}{\pi}
    \int \frac{d^3k}{2 \left(k^2+\lambda^2a^2\right) }
         \frac{-1}{ \left[ k + \sum_j \tilde{v}_j  k_j \right] }
  = \frac{g^2 C_F}{16\pi^2}
    \frac{-2}{\tilde{v}} \ln\!\left(\frac{1+\tilde{v}}{1-\tilde{v}}\right)
    \ln\!\left(\frac{4R^2}{\lambda^2a^2}\right).
    \label{eq:L}
\end{eqnarray}
However, this gives a $\lambda$-dependent $\delta(\tilde{v})$.  In spherical
coordinates, $\delta(\tilde{v})$ has the form
\begin{equation}
\delta(\tilde{v})
  =   \frac{g^2 C_F}{16\pi^2} \frac{2}{\tilde{v}}
      \int \frac{k\,dk}{k^2+\lambda^2a^2}
           \left[ \ln\!\left(\frac{\sqrt{k^2+\lambda^2a^2}+\tilde{v} k}
                                  {\sqrt{k^2+\lambda^2a^2}-\tilde{v} k}\right)
                - \ln\!\left(\frac{1+\tilde{v}}{1-\tilde{v}}\right) \right].
\end{equation}
So long as $\tilde{v}$ is finite, we can take the limit as $\lambda \rightarrow
0$.  However, if we want both $\tilde{v} \rightarrow 0$ and $\lambda
\rightarrow 0$, a problem arises: the result in the limit $\lambda \rightarrow
0$ is not the result at $\lambda = 0$.  This is a case in which the limits
cannot be interchanged.  To be rigorous, we break up the integration into a
region for $k < \lambda$ and for $k > \lambda$.
\begin{eqnarray}
\delta(\tilde{v} \rightarrow 0,\lambda \rightarrow 0 )
& = & \frac{g^2 C_F}{16\pi^2} 4
      \left[ \lim_{\epsilon \rightarrow 0} \int_\epsilon^\lambda dk
             \left( \frac{k^2}{\left(k^2+\lambda^2a^2
\right)^{{}^{3}\!\!/\!_{2}}}
                  - \frac{k}{\left(k^2+\lambda^2a^2\right)} \right)
      \right. \nonumber \\ & & \left.
           + \int_\lambda^R dk
             \left( \frac{k^2}{\left(k^2+\lambda^2a^2
\right)^{{}^{3}\!\!/\!_{2}}}
                  - \frac{k}{\left(k^2+\lambda^2a^2\right)} \right)
             \right]  \\
& = & \frac{g^2 C_F}{16\pi^2} 4 \lim_{\lambda \rightarrow 0}
      \left[-1 + \frac{1}{2} \ln\!\left(4\frac{R^2}{\lambda^2a^2}\right)
           - \frac{1}{4} \ln\!\left(\frac{R^2}{\lambda^2a^2}\right) \right]
      \nonumber \\ & &
    + \lim_{\lambda \rightarrow 0} O\!( {}^{\lambda^2a^2}\!\!/\!_{R^2} )
    + \lim_{\lambda \gg \epsilon \rightarrow 0}
      O\!( {}^{\epsilon^3}\!\!/\!_{\lambda^3} )
      \label{eq:blows-up}  \\
& = & \infty
\end{eqnarray}
While a $\lambda$ divergence was expected for $\delta V$, the difference
$\delta(\tilde{v})$ must be finite.  $\delta(\tilde{v} \rightarrow 0,\lambda
\rightarrow 0 )$ is infinite because the logarithms do not cancel exactly.  It
happens that $\delta(\lambda)$ has a minimum around $\lambda\approx10^{-5}$; at
this value, if $\tilde{v}$ is taken to zero, then Aglietti's
$d(\tilde{v}=0,\tilde{v}'=0) = -4.53$ can be calculated from $\lim_{\tilde{v}
\rightarrow 0} \left[ \, \delta V(\tilde{v},0) - \delta V^c(\tilde{v},0) +
\mbox{finite part of }V^{cc}(\tilde{v},0) \right] = -5.75$ and
$\delta(\tilde{v} \rightarrow 0, \lambda \approx 10^{-5}) = -1.22$.  However,
Eq.~(\ref{eq:blows-up}) clearly shows that $\delta(\lambda \rightarrow 0)$
blows up.  This can be seen for $\lambda < 10^{-5}$.

To avoid this problem, we write Eq.~(\ref{eq:deltaVc0}) as
\begin{eqnarray}
\delta V^{c}\!\left(\tilde{v},\tilde{v}'=0\right)
& = & \frac{g^2 C_F}{16\pi^2} \frac{2}{\pi}
      \int \frac{d^3k}{2 \left(k^2+\lambda^2a^2\right)^{{}^{3}\!\!/\!_{2}}}
           \frac{-1}
                { \left[ 1 + \frac{\vec{\tilde{v}} \cdot \vec{k}}
                                  {\sqrt{k^2+\lambda^2a^2}}
                         \right] },
\end{eqnarray}
which expands as follows and allows a better definition of $\delta
V^{cc}\!\left(\tilde{v},\tilde{v}'=0\right)$:
\begin{eqnarray}
\delta V^{c}\!\left(\tilde{v},\tilde{v}'=0\right)
& \approx & \frac{g^2 C_F}{16\pi^2} \frac{2}{\pi}
      \int \frac{d^3k}{2 \left(k^2+\lambda^2a^2\right)^{{}^{3}\!\!/\!_{2}}}
           \frac{-1}
                { \left[ 1 + \vec{\tilde{v}} \cdot \hat{k}
                             \left(1-\frac{1}{2} \frac{\lambda^2a^2}{k^2}
                                    \right) \right] } \\
\delta V^{cc}\!\left(\tilde{v},\tilde{v}'=0\right)
& = & \frac{g^2 C_F}{16\pi^2} \frac{2}{\pi}
      \int \frac{d^3k}{2 \left(k^2+\lambda^2a^2\right)^{{}^{3}\!\!/\!_{2}}}
           \frac{-1}
                { \left[ 1 + \vec{\tilde{v}} \cdot \hat{k} \right] }.
\end{eqnarray}
We find that this makes $\delta'(\tilde{v})$ stable to small $\lambda$ and that
it generalizes to give useful results when $\tilde{v}' \ne 0$.  In the equation
analogous to Eq.~(\ref{eq:blows-up}), the logarithms cancel and the result is
finite in the $\lambda \rightarrow 0$ limit.


\subsubsection{Vertex Correction with $\tilde{v}' \ne 0$}

This case requires the continuum-like expression for Eq.~(\ref{eq:deltaV}).
Recall Eq.~(\ref{eq:deltaVc}).  Again there are problems if we use Aglietti's
trick of setting $\lambda$ to zero in the factors with $\tilde{v}$ and
$\tilde{v}'$.  The problems are: (1) the $\lambda$-dependence is incorrect
(which implies that the difference is $\lambda$-dependent), (2) the limit as
$\tilde{v}' \rightarrow 0$ does not reproduce the results of the previous
subsection, and (3) the integral is too difficult.  So, once again, we will try
to retain the $\lambda$-dependence as follows: we approximate
\begin{eqnarray}
\delta V^{c}\!\left(\tilde{v},\tilde{v}'\right)
& = & \frac{g^2 C_F}{16\pi^2} \frac{2}{\pi}
      \int \frac{d^3k}{2 \left(k^2+\lambda^2a^2 \right)^{{}^{3}\!\!/\!_{2}}}
           \frac{-1 + \sum_j \tilde{v}_j \tilde{v}_j' }
                { \left[ 1
                       + \frac{ \vec{\tilde{v}} \cdot \vec{k} }
                              { \sqrt{k^2+\lambda^2a^2}  } \right]
                  \left[ 1
                       + \frac{ \vec{\tilde{v}'} \cdot \vec{k} }
                              {  \sqrt{k^2+\lambda^2a^2}  } \right] }
      \nonumber
\end{eqnarray}
by
\begin{eqnarray}
\delta V^{cc}\!\left(\tilde{v},\tilde{v}'\right)
& = & \frac{g^2 C_F}{16\pi^2} \frac{2}{\pi}
      \int \frac{d^3k}{2 \left(k^2+\lambda^2a^2 \right)^{{}^{3}\!\!/\!_{2}}}
           \frac{-1 + \sum_j \tilde{v}_j \tilde{v}_j' }
                { \left[ 1 + \vec{\tilde{v}} \cdot \hat{k} \right]
                  \left[ 1 + \vec{\tilde{v}'} \cdot \hat{k} \right] }.
      \nonumber
\end{eqnarray}
While this does solve both the $\lambda$-dependence problem and the $\tilde{v}'
\rightarrow 0$ problem, it only barely solves the difficulty of the integral.
However, in spherical coordinates, it allows the $|k|$ integral to be separated
from the angular integration.  We can solve this integral by doing the $|k|$
integration analytically. (Since this is where the $\lambda$-divergence exists,
it is the only piece that needs to be done analytically anyway.)  Having thus
removed the $\ln(\lambda)$ term, we can numerically calculate and add back the
angular integration along with the finite piece of the $|k|$ integration.

Since this is symmetric in $\tilde{v}$ and $\tilde{v}'$, the results for the
vertex correction should be also.  Our extension to nonzero $\tilde{v}'$ shows
that this is the case: Tables~\ref{t:my-d-backward} and~\ref{t:my-d-forward}
show our results for backward and forward difference vertex correction at
general velocities.  Notice that the results are symmetric about the diagonal,
$\tilde{v}=\tilde{v}'$.
\begin{table}
\caption[Backward Difference Vertex Correction for General Velocities]
        {\label{t:my-d-backward}
         The finite piece of the backward difference vertex correction
         $d(\tilde{v},\tilde{v}')$ for $v_x = v_y = v'_x = v'_y = 0$.}
\begin{center}
\begin{tabular}{cccccccc} \hline \hline
${}_{\tilde{v}'}\backslash^{\tilde{v}}$ & 0.0 & 0.1 & 0.2 & 0.3 & 0.4 & 0.5 &
0.6
\\ \hline \hline
0.0 & $-4.528(2)$ & $-4.579(2)$ & $-4.756(2)$ & $-5.088(3)$
    & $-5.642(3)$ & $-6.592(4)$ & $-8.454(5)$  \\
0.1 & $-4.583(2)$ & $-4.459(2)$ & $-4.446(2)$ & $-4.567(3)$
    & $-4.892(3)$ & $-5.562(3)$ & $-7.018(4)$ \\
0.2 & $-4.755(2)$ & $-4.447(2)$ & $-4.228(3)$ & $-4.131(3)$
    & $-4.186(3)$ & $-4.523(3)$ & $-5.525(4)$  \\
0.3 & $-5.088(3)$ & $-4.570(3)$ & $-4.126(3)$ & $-3.768(3)$
    & $-3.514(3)$ & $-3.457(4)$ & $-3.890(4)$ \\
0.4 & $-5.640(3)$ & $-4.890(3)$ & $-4.183(3)$ & $-3.517(3)$
    & $-2.882(4)$ & $-2.320(4)$ & $-2.007(5)$  \\
0.5 & $-6.598(4)$ & $-5.556(3)$ & $-4.523(4)$ & $-3.460(4)$
    & $-2.320(4)$ & $-1.071(5)$ & $0.283(6)$  \\
0.6 & $-8.452(4)$ & $-7.022(4)$ & $-5.529(5)$ & $-3.884(4)$
    & $-2.003(5)$ & $0.280(6)$ & $3.322(8)$  \\
\hline\hline
\end{tabular}
%
%
\end{center}
%
%
%
\caption[Forward Difference Vertex Correction for General Velocities]
        {\label{t:my-d-forward}
         The finite piece of the forward difference vertex correction
         $d(\tilde{v},\tilde{v}')$ for $v_x = v_y = v'_x = v'_y = 0$.}
\begin{center}
\begin{tabular}{cccccccc} \hline \hline
${}_{\tilde{v}'}\backslash^{\tilde{v}}$ & 0.0 & 0.1 & 0.2 & 0.3 & 0.4 & 0.5 &
0.6
\\ \hline \hline
0.0 & $-4.528(2)$ & $-4.513(2)$ & $-4.471(2)$ & $-4.402(3)$
    & $-4.280(3)$ & $-4.102(3)$ & $-3.822(4)$  \\
0.1 & $-4.516(2)$ & $-4.527(2)$ & $-4.509(2)$ & $-4.462(3)$
    & $-4.370(3)$ & $-4.221(3)$ & $-3.971(4)$  \\
0.2 & $-4.471(2)$ & $-4.508(2)$ & $-4.513(3)$ & $-4.487(3)$
    & $-4.437(3)$ & $-4.315(3)$ & $-4.108(4)$  \\
0.3 & $-4.402(3)$ & $-4.457(3)$ & $-4.494(3)$ & $-4.505(3)$
    & $-4.472(3)$ & $-4.396(3)$ & $-4.224(4)$ \\
0.4 & $-4.283(3)$ & $-4.371(3)$ & $-4.431(3)$ & $-4.474(3)$
    & $-4.481(3)$ & $-4.446(4)$ & $-4.325(4)$  \\
0.5 & $-4.108(4)$ & $-4.219(3)$ & $-4.316(3)$ & $-4.398(4)$
    & $-4.443(4)$ & $-4.459(4)$ & $-4.402(4)$  \\
0.6 & $-3.822(4)$ & $-3.973(4)$ & $-4.112(4)$ & $-4.226(4)$
    & $-4.320(4)$ & $-4.398(4)$ & $-4.424(4)$  \\
\hline\hline
\end{tabular}
%
%
\end{center}
\end{table}
Notice also that the first row and the first column of
Table~\ref{t:my-d-backward} both reproduce the backward-difference results of
$\tilde{v}'=0$ in Table~\ref{t:my-ag2}.  Table~\ref{t:my-d-forward} shows the
results for the forward-difference vertex correction at general velocities and
the first row and column reproduce the forward-difference results of
Table~\ref{t:my-ag2}.


\section{Lattice to Continuum Matching}
\label{s:matching}

To renormalize the Isgur-Wise function, which is proportional to the current in
Eq.~(\ref{eq:current}), we need the current renormalization, which can be
assembled from the wavefunction and vertex renormalizations calculated in the
previous section.  This involves
\begin{eqnarray}
Z_Q^{{}^{1}\!\!/\!_{2}}(v) Z_V(v,v') Z_Q^{{}^{1}\!\!/\!_{2}}(v')
& = & \left[1+\frac{1}{2}\delta Z_Q(v)\right]
      \left[1+\phantom{\frac{1}{1}\!\!\!\!}\delta V(v,v')\right]
      \left[1+\frac{1}{2}\delta Z_Q(v')\right]
      \label{eq:latt-renorm} \\
& = & \left\{ 1
            + \frac{1}{2} \left[ \delta Z_Q(v) + \delta Z_Q(v')\right]
            + \delta V(v,v') \right\}.
      \nonumber
\end{eqnarray}
So, following Aglietti's lead~\cite{Aglietti94}, we define
\begin{eqnarray}
f(\tilde{v},\tilde{v}') & = & {}^{\underline 1}_{2} \left[ e(\tilde{v}) +
e(\tilde{v}') \right]
            + d(\tilde{v},\tilde{v}') \nonumber \\
f'(\tilde{v},\tilde{v}') & = & {}^{1}_{\overline 2} \left[ e'(\tilde{v}) +
e'(\tilde{v}') \right]
             + d(\tilde{v},\tilde{v}') \nonumber
\end{eqnarray}
where a reduced Isgur-Wise correction, $f'$, is defined using the reduced
wavefunction, $e'$, which was used with a fit model of the form $e^{-mt}$.
Since the wavefunction reduction does not affect the vertex correction, $d$,
the perturbative factor for the lattice Isgur-Wise function is:
\begin{equation}
Z_{\xi}^{\rm l} = 1 +
\frac{g^2}{12\pi^2}
\left\{ -2 \left[ 1 - (v \cdot v') \; r(v \cdot v') \right]
       \ln (\lambda a)^2
     + f'(\tilde{v},\tilde{v}') \right\}
\label{eq:latt renorm}
\end{equation}
where the divergences have been isolated to calculate the finite pieces and
$r(v \cdot v')$ has been defined by Eq.~(\ref{eq:rfalk}).  If we did not wish
to use the reduced value, the divergence would stay the same and we would
merely replace the $f'$ with $f$.  We have already shown
(Table~\ref{t:lattice-falk}) that the divergent piece of the lattice vertex
correction cancels exactly with that of the continuum; thus the lattice
logarithm coefficient is written with the same form as for the continuum
correction, Eq.~(\ref{eq:cont renorm}).  Now the continuum correction,
$Z_{\xi}^{\rm c}$, can be divided by the lattice correction, $Z_{\xi}^{\rm l}$,
to find the lattice to continuum matching factor
\begin{equation}
Z_{\xi}^{\rm cl}(v,v') =  1 +
\frac{g^2}{12\pi^2}
\left\{ 2 \left[ 1 - (v \cdot v') \; r(v \cdot v') \right]
       \ln (\mu a)^2 - f'(\tilde{v},\tilde{v}') \right\}.
\label{eq:Ztwothree}
\end{equation}
The expression in Eq.~(\ref{eq:Ztwothree}) is suitable for renormalizing the
Isgur-Wise function extracted by taking ratios of two and three point
functions~\cite{Bowler:1995bp}.  However, to improve statistics, HQET
simulations extract the Isgur-Wise function using ratios of three point
functions only~\cite{Christensen97b,Hashimoto96,Mandula94a}.  We discuss this
additional complication below.

Our results for $d$, $f$, and $f'$ are listed in the following tables.  Recall
Tables~\ref{t:my-d-backward} and~\ref{t:my-d-forward} show our results for
backward and forward difference vertex correction at general velocities.
Table~\ref{t:my-ag2} lists our results for the vertex correction, the current
correction, and the reduced current correction in the backward and forward
difference actions for $\tilde{v}'=0$.  The backward difference reproduces
Aglietti's results.
\begin{table}
\caption[Vertex and Current Corrections for $\tilde{v}'=0$]
        {The finite piece of the vertex correction,
         $d(\tilde{v},\tilde{v}')$, of the forward- and
         backward-difference actions for $v_x = v_y = 0$ and $\vec
         v'=\vec 0$.  The backward-difference action results should and
         do reproduce Aglietti's table up to the correction mentioned
         in the text of this paper.  Also listed is the current
         correction $f(\tilde{v},\tilde{v}') = \frac{1}{2}
         e(\tilde{v}) + \frac{1}{2} e(\tilde{v}') +
         d(\tilde{v},\tilde{v}')$ and the reduced current correction
         $f'(\tilde{v},\tilde{v}') = \frac{1}{2} e'(\tilde{v}) +
         \frac{1}{2} e'(\tilde{v}') + d(\tilde{v},\tilde{v}')$
         which form the correction for the lattice Isgur-Wise
         function.}
\begin{center}
\begin{tabular}{ccccccc} \hline \hline
& \multicolumn{3}{c}{Backward Difference} & \multicolumn{3}{c}{Forward
Difference} \\
$\tilde{v}$ & $d(\tilde{v},\tilde{v}'=0)$ & $f(\tilde{v},\tilde{v}'=0)$ &
$f'(\tilde{v},\tilde{v}'=0)$ & $d(\tilde{v},\tilde{v}'=0)$ &
$f(\tilde{v},\tilde{v}'=0)$ & $f'(\tilde{v},\tilde{v}'=0)$ \\ \hline
0.0 &\ -4.526(2) & 19.92(1) & 0.000(2) & -4.527(2) & -19.94(1) &\ \ 0.000(2)\\
0.1 &\ -4.578(2) & 19.96(1) & 0.122(2) & -4.511(2) & -20.09(1) &\ -0.174(2)\\
0.2 &\ -4.757(2) & 20.08(1) & 0.489(2) & -4.474(2) & -20.58(1) &\ -0.740(2) \\
0.3 &\ -5.089(2) & 20.33(2) & 1.129(2) & -4.401(2) & -21.50(1) &\ -1.755(2) \\
0.4 &\ -5.639(4) & 20.63(2) & 2.100(4) & -4.282(2) & -22.93(1) &\ -3.364(2) \\
0.5 &\ -6.597(4) & 21.09(2) & 3.459(4) & -4.104(4) & -25.21(1) &\ -5.844(2) \\
0.6 &\ -8.432(6) & 21.50(2) & 5.23(1)\ & -3.800(4) & -28.89(1) &\ -9.755(4) \\
0.7 & -14.80(1)\ & 19.53(4) & 5.57(2)\ & -3.354(6) & -35.32(2) & -16.529(8)
\\ \hline \hline
\end{tabular}
\label{t:my-ag2}
%
%
\end{center}
\end{table}
Tables~\ref{t:my-f-backward} and~\ref{t:my-fp-backward} show our results for
backward difference current and reduced current corrections at general
velocities.  Again, the results are symmetric about the diagonal.
\begin{table}
\caption[Backward Difference Current Correction for General Velocities]
        {\label{t:my-f-backward}
         The finite piece of the backward difference current
         correction, $f(\tilde{v},\tilde{v}')$, for $v_x = v_y = 0$ and
         $v'_x = v'_y = 0$.}
\begin{center}
\begin{tabular}{cccccccc} \hline \hline
${}_{\tilde{v}'}\backslash^{\tilde{v}}$ & 0.0 & 0.1 & 0.2 & 0.3 & 0.4 & 0.5 &
0.6
\\ \hline \hline
0.0 & $19.903(8)$ & $19.958(8)$ & $20.100(9)$ & $20.324(10)$
    & $20.654(9)$ & $21.093(9)$ & $21.51(1)$ \\
0.1 & $19.961(8)$ & $20.195(10)$ & $20.485(8)$ & $20.921(10)$
    & $21.516(10)$ & $22.245(10)$ & $23.03(1)$  \\
0.2 & $20.087(8)$ & $20.498(9)$ & $21.022(9)$ & $21.662(9)$
    & $22.522(9)$ & $23.565(10)$ & $24.84(1)$  \\
0.3 & $20.320(9)$ & $20.947(9)$ & $21.681(9)$ & $22.57(1)$
    & $23.73(1)$ & $25.20(1)$ & $27.02(1)$ \\
0.4 & $20.646(10)$ & $21.494(9)$ & $22.495(9)$ & $23.74(1)$
    & $25.25(1)$ & $27.24(1)$ & $29.78(1)$  \\
0.5 & $21.090(9)$ & $22.228(9)$ & $23.579(10)$ & $25.175(10)$
    & $27.20(1)$ & $29.86(1)$ & $33.48(1)$  \\
0.6 & $21.499(10)$ & $23.017(10)$ & $24.84(1)$ & $27.02(1)$
    & $29.75(1)$ & $33.47(1)$ & $38.75(1)$  \\
\hline\hline
\end{tabular}
\end{center}
%
%
%
\vfill
%
%
\caption[Backward Difference Reduced Current Correction for General
         Velocities]
        {\label{t:my-fp-backward}
         The finite piece of the backward difference reduced current
         correction, $f'(\tilde{v},\tilde{v}')$, for $v_x = v_y = 0$
         and $v'_x = v'_y = 0$.}
\begin{center}
%
%
\begin{tabular}{ccccccccc} \hline \hline
${}_{\tilde{v}'}\backslash^{\tilde{v}}$ & 0.0 & 0.1 & 0.2 & 0.3 & 0.4 & 0.5 &
0.6
\\ \hline \hline
0.0 & $0.004(3)$ & $0.118(2)$ & $0.491(2)$ & $1.134(3)$
    & $2.098(3)$ & $3.458(4)$ & $5.228(6)$ \\
0.1 & $0.119(2)$ & $0.422(3)$ & $0.975(2)$ & $1.821(3)$
    & $3.021(3)$ & $4.675(4)$ & $6.822(6)$ \\
0.2 & $0.488(2)$ & $0.974(2)$ & $1.733(3)$ & $2.810(3)$
    & $4.273(4)$ & $6.258(5)$ & $8.858(6)$ \\
0.3 & $1.128(3)$ & $1.822(3)$ & $2.812(3)$ & $4.142(4)$
    & $5.922(4)$ & $8.292(5)$ & $11.484(7)$ \\
0.4 & $2.096(3)$ & $3.026(4)$ & $4.274(4)$ & $5.922(4)$
    & $8.073(5)$ & $10.954(6)$ & $14.875(7)$ \\
0.5 & $3.459(4)$ & $4.672(4)$ & $6.250(4)$ & $8.291(5)$
    & $10.957(6)$ & $14.504(7)$ & $19.477(9)$ \\
0.6 & $5.219(6)$ & $6.839(6)$ & $8.856(6)$ & $11.477(7)$
    & $14.893(7)$ & $19.470(9)$ & $26.10(1)$ \\
\hline\hline
\end{tabular}
%
%
\end{center}
\end{table}
Tables~\ref{t:my-f-forward} and~\ref{t:my-fp-forward} for the forward
difference current and reduced current corrections at general velocities are
also symmetric about the diagonal.  Notice that the first rows and columns of
Tables~\ref{t:my-d-backward}, \ref{t:my-d-forward}, and~\ref{t:my-f-backward}
through~\ref{t:my-fp-forward} reproduce Table~\ref{t:my-ag2}.  Notice also that
although the different actions give the same result in the static limit ($v
\rightarrow 0$, $v' \rightarrow 0$), this is not the case at any other
velocity.
\begin{table}
\caption[Forward Difference Current Correction for General Velocities]
        {\label{t:my-f-forward}
         The negative of the finite piece of the forward difference
         current correction, $-f(\tilde{v},\tilde{v}')$, for \mbox{$v_x
         = v_y = 0$} and $v'_x = v'_y = 0$.}
\begin{center}
\begin{tabular}{cccccccc} \hline \hline
${}_{\tilde{v}'}\backslash^{\tilde{v}}$ & 0.0 & 0.1 & 0.2 & 0.3 & 0.4 & 0.5 &
0.6
\\ \hline \hline
0.0 & $19.923(6)$ & $20.086(6)$ & $20.591(7)$ & $21.500(7)$
    & $22.953(7)$ & $25.223(8)$ & $28.897(9)$ \\
0.1 & $20.095(7)$ & $20.282(7)$ & $20.786(7)$ & $21.714(7)$
    & $23.206(7)$ & $25.513(8)$ & $29.217(8)$  \\
0.2 & $20.591(7)$ & $20.784(7)$ & $21.342(7)$ & $22.297(7)$
    & $23.810(7)$ & $26.136(8)$ & $29.882(10)$ \\
0.3 & $21.491(7)$ & $21.742(7)$ & $22.298(7)$ & $23.268(8)$
    & $24.819(8)$ & $27.188(9)$ & $30.987(10)$ \\
0.4 & $22.941(8)$ & $23.192(8)$ & $23.794(7)$ & $24.824(8)$
    & $26.394(9)$ & $28.829(9)$ & $32.66(1)$ \\
0.5 & $25.220(8)$ & $25.498(9)$ & $26.138(8)$ & $27.177(8)$
    & $28.804(9)$ & $31.283(10)$ & $35.15(1)$ \\
0.6 & $28.885(9)$ & $29.22(1)$ & $29.897(9)$ & $30.993(10)$
    & $32.64(1)$ & $35.16(1)$ & $39.16(1)$  \\
0.7 & $35.32(1)$ & $35.68(1)$ & $36.41(1)$ & $37.59(1)$
    & $39.30(1)$ & $41.90(1)$ & $46.01(2)$ \\
\hline\hline
\end{tabular}
\end{center}
%
%
%
\vfill
%
%
\caption[Forward Difference Reduced Current Correction for General
         Velocities]
        {\label{t:my-fp-forward}
         The negative of the finite piece of the forward difference
         reduced current correction, $-f'(\tilde{v},\tilde{v}')$, for
         $v_x = v_y = 0$ and $v'_x = v'_y = 0$.}
\begin{center}
%
%
\begin{tabular}{cccccccc} \hline \hline
${}_{\tilde{v}'}\backslash^{\tilde{v}}$ & 0.0 & 0.1 & 0.2 & 0.3 & 0.4 & 0.5 &
0.6
\\ \hline \hline
0.0 & $-0.004(3)$ & $0.178(2)$ & $0.737(2)$ & $1.755(2)$
    & $3.365(3)$ & $5.843(3)$ & $9.773(4)$  \\
0.1 & $0.181(2)$ & $0.380(2)$ & $0.968(2)$ & $2.008(2)$
    & $3.645(3)$ & $6.154(3)$ & $10.113(4)$ \\
0.2 & $0.740(2)$ & $0.967(2)$ & $1.583(2)$ & $2.649(3)$
    & $4.311(3)$ & $6.849(3)$ & $10.851(4)$ \\
0.3 & $1.759(2)$ & $2.009(3)$ & $2.652(3)$ & $3.742(3)$
    & $5.441(3)$ & $8.020(4)$ & $12.061(5)$ \\
0.4 & $3.366(3)$ & $3.642(3)$ & $4.312(3)$ & $5.442(3)$
    & $7.173(3)$ & $9.796(4)$ & $13.881(5)$ \\
0.5 & $5.841(3)$ & $6.154(3)$ & $6.858(3)$ & $8.022(4)$
    & $9.796(4)$ & $12.450(5)$ & $16.618(6)$ \\
0.6 & $9.772(4)$ & $10.118(4)$ & $10.855(4)$ & $12.059(5)$
    & $13.886(5)$ & $16.610(6)$ & $20.845(8)$ \\
0.7 & $16.525(6)$ & $16.911(6)$ & $17.687(6)$ & $18.964(7)$
    & $20.839(8)$ & $23.675(9)$ & $27.996(10)$ \\
\hline\hline
\end{tabular}
%
%
\end{center}
\end{table}

For continuum HQET in the $\overline{MS}$ renormalization scheme at zero recoil
($v \cdot v'=1$), $Z_\xi^{\rm c} = 1$ and the finite piece is
zero~\cite{Neubert94a}.  This corresponds to the diagonal ($v=v'$) of the
tables which contain our results.  On the lattice, however, if the conserved
current is not used, $f'(v,v)$ is not constrained to be zero.  We account for
this next.


To deal with the finite piece of the renormalization, we note that the numeric
extraction of the Isgur-Wise function on the lattice does not calculate the
Isgur-Wise function directly.  The numerical extraction is more manageable
using the technique of Mandula and Ogilvie~\cite{Mandula94a} where the ratio of
the three-point quark propagator, $G$ (defined explicitly in our concurrent
numerical paper~\cite{Christensen98}), gives a ratio of lattice Isgur-Wise
functions:
\begin{eqnarray}
\label{eq:numerical}
   \frac{   4 v_0 v'_0   }
        { (v_0 + v'_0)^2 }
\frac{ G^{v,v'}(  \tau  ) G^{v',v }( \tau ) }
     { G^{v,v }(  \tau  ) G^{v',v'}( \tau ) }
& \ \begin{array}{c}\\^{\tau \gg 1}\end{array}
                        \!\!\!\!\!\!\!\!\!\!\!\!\!\!\!
                        -\!\!\!-\!\!\!\longrightarrow &
   \frac{\left| \xi (v,v') \right|\,\left| \xi (v',v ) \right|}
        {\left| \xi (v,v ) \right|\,\left| \xi (v',v') \right|}
\end{eqnarray}
This technique exploits the continuum normalization of the Isgur-Wise function
at zero recoil
\begin{equation}
\label{eq:iwone}
\xi(v \cdot v) = \xi(1) = 1.
\end{equation}
Since $v_\mu^2$ is normalized to $1$, the denominator of
Eq.~(\ref{eq:numerical}) can be set to unity in the continuum.
This ratio also allows the normalizations and smearing-function dependence to
cancel, so we expect that
\begin{eqnarray}
\frac{\left| Z_\xi^{\rm cl}(v ,v') \xi^{\rm latt} (v ,v') \right|
      \left| Z_\xi^{\rm cl}(v',v ) \xi^{\rm latt} (v',v ) \right| }
     {\left| Z_\xi^{\rm cl}(v ,v ) \xi^{\rm latt} (v ,v ) \right| \,
      \left| Z_\xi^{\rm cl}(v',v') \xi^{\rm latt} (v',v') \right|  }
& \ \ \begin{array}{c}\\^{a \rightarrow 0\ }\end{array}
                        \!\!\!\!\!\!\!\!\!\!\!\!\!\!\!\!\!\!\!
                        -\!\!\!-\!\!\!-\!\!\!\longrightarrow
& \frac{\left| \xi^{\rm cont} (v \cdot v') \right|^2}
       {\left| \xi^{\rm cont} (  1  ) \right|^2}
 = \left| \xi^{\rm cont} (v \cdot v') \right|^2.
 \nonumber
\end{eqnarray}
Thus, our unrenormalized calculation of
\[ \xi_{\rm ratio}(v,v') \equiv \left( \frac{\xi^{\rm latt}(v,v') \xi^{\rm
latt}(v',v )}
                                        {\xi^{\rm latt}(v,v ) \xi^{\rm
latt}(v',v')}
                           \right)^{{}^{1}\!\!/\!_{2}}  \]
must be renormalized by
\begin{equation}  \label{eq:superz}
\mbox{$Z_{\rm ratio}^{\rm cl}$}(v ,v')
 = \left( \frac{Z_\xi^{\rm cl}(v ,v') Z_\xi^{\rm cl}(v',v )}
               {Z_\xi^{\rm cl}(v ,v ) Z_\xi^{\rm cl}(v',v')}
          \right)^{{}^{1}\!\!/\!_{2}}
\end{equation}
written as
\begin{equation}
\mbox{$Z_{\rm ratio}^{\rm cl}$}(v,v') \, \xi_{\rm ratio}(v,v')
\ \begin{array}{c}\\^{a \rightarrow 0}\end{array}
                        \!\!\!\!\!\!\!\!\!\!\!\!\!\!\!
                        -\!\!\!-\!\!\!\longrightarrow
  \xi^{\rm cont}(v \cdot v').
  \label{eq:xi-renorm}
\end{equation}

On the lattice, $\xi^{\rm latt}(v, v)$ does not obey Eq.~(\ref{eq:iwone})
unless a conserved current is used; nevertheless, $\xi_{\rm ratio}(v,v')$ (by
definition) acts like the continuum Isgur-Wise function even if the conserved
current is not used.  Without the conserved current, $\xi^{\rm latt}(v,v) \neq
1$, but the normalization cancels in the ratio so that $\xi_{\rm
ratio}(v,v)=1$.  Thus, $Z_{\rm ratio}^{\rm cl}$ will be symmetric in $v$ and
$v'$ and will have the property $Z_{\rm ratio}^{\rm cl}(v,v)=1$.


Expanding Eq.~(\ref{eq:superz}), we find
\begin{eqnarray}
\mbox{$Z_{\rm ratio}^{\rm cl}$}
  =   1
& + & \frac{1}{2} \frac{g^2 C_F}{16\pi^2}
      \left\{ 2 \left[ \left( 1 -(v \cdot v')\;r\!\left(v \cdot v'\right)
\right)
                     + \left( 1 -(v'\cdot v )\;r\!\left(v'\cdot v \right)
\right)
                       \right.\right.\nonumber\\&&\left.\left.
                     - \left( 1 -(v \cdot v )\;r\!\left(v \cdot v \right)
\right)
                     - \left( 1 -(v'\cdot v')\;r\!\left(v'\cdot v'\right)
\right)
                       \right] \ln\!\left(\mu a\right)^2
              \right.\nonumber\\&&\left.
            - f'(\tilde{v},\tilde{v}') - f'(\tilde{v}',\tilde{v}) +
f'(\tilde{v},\tilde{v}) + f'(\tilde{v}',\tilde{v}') \right\}.
\end{eqnarray}
Using $v \cdot v = v' \cdot v' = r(1) = 1$ and
$f'(\tilde{v},\tilde{v}')=f'(\tilde{v}',\tilde{v})$, this reduces to
\begin{equation}
Z_{\rm ratio}^{\rm cl}(v,v') =  1 +
\frac{g^2}{12\pi^2}
\left[ 2 \left( 1 - (v \cdot v') \; r(v \cdot v') \right)
       \ln (\mu a)^2
     - f'(\tilde{v},\tilde{v}') +  \frac{f'(\tilde{v},\tilde{v}) +
f'(\tilde{v}',\tilde{v}')}{2} \right].
\label{eq:z-match}
\end{equation}
which not only has the correct divergent coefficient but we also see a new
finite piece which is manifestly zero on the diagonal.  The wavefunction
renormalization cancels explicitly in Eq.~(\ref{eq:z-match}), so $f'$ can be
replaced by the vertex correction $d$.


\section{Conclusions}
\label{s:conclusion}

We have calculated the renormalization of the lattice $b \rightarrow c$ current
by considering the lattice Isgur-Wise function.  This calculation extends
previous work by including tadpole improvement, by extending to non-zero
initial and final velocities, and by considering forward as well as backward
difference actions.

By considering the forward difference action and the backward difference action
side-by-side, we find non-trivial differences between the two.  The practical
difference in a lattice calculation is that the backward difference requires a
matrix inversion at each step of the calculation.  The differences in the
renormalization are that the gluon poles over which one integrates are
interchanged; away from the static limit, the reduced values are no longer
equal; and the velocity renormalization, when expanded as powers of the
velocity, stay small for the forward difference, but grow large for the
backward difference.

Of greater concern is that the velocity renormalization is not terribly small.
We have shown that the velocity renormalization can be expanded in small
velocity and that the coefficients remain on the order of unity at higher
orders (at least for the forward difference action).  These coefficients are
given here to $O\!( v^6 )$.  The nonperturbative calculations are giving
smaller renormalizations~\cite{Mandula:1997hb,Hashimoto96} and these should be,
in principle, more reliable. This should be considered in more detail,
especially the slow HQET for the forward difference action.

Although our results confirm other groups' calculations where they overlap, the
integrals and divergences are subtle and must be managed with care.  When we
combine our renormalizations into a current correction with the ratio
introduced by Mandula and Ogilvie~\cite{Mandula94a}, such that the finite piece
of the current correction is $- f'(\tilde{v},\tilde{v}') + \frac{1}{2} \left[
f'(\tilde{v},\tilde{v}) + f'(\tilde{v}',\tilde{v}') \right]$, we find that all
of our results have the appropriate limits and cancelations.  These expressions
are used in our concurrent numerical paper~\cite{Christensen98} to compute the
slope of the Isgur-Wise function using lattice HQET.

%

\appendix
\renewcommand{\theequation}{\Alph{section}\arabic{equation}}
\setcounter{equation}{0}


\section{Appendix: Tadpole Improvement}
\label{a:tadpole}

Tadpole improvement is a mean field improvement~\cite{Lepage93} which (at
lowest order) cancels the effects of the large ``tadpole'' Feynman diagrams.
In the HQET, there is no coefficient (analogous to $\kappa$ in the Wilson
action) which is common to both $U_t$ and $U_j$ and which allows one to {\it a
posteriori\/} tadpole improve any previous calculation which was not tadpole
improved.  Fortunately, as noticed by Mandula and
Ogilvie~\cite{Mandula:1997hb}, the evolution equation can be written such that
the $u_0$ is grouped with $\tilde{v}_j = {}^{v_j}\!\!/\!_{v_0}$.  Thus,
tadpole-improved ($tad$) Monte-Carlo data can be {\it constructed\/} from the
non-tadpole-improved ($nt$) data by replacing $v^{\rm nt} \rightarrow v^{\rm
tad}$ and by including two overall multiplicative factors (${}^{v_0^{\rm
nt}}\!\!/\!_{v_0^{\rm tad}}$ was not included by Mandula and Ogilvie):
\begin{eqnarray}
G^{\rm tad}(t;\tilde{v}^{\rm tad}, v_0^{\rm tad})
& \!\!=\!\! & u_0^{-t} \frac{v_0^{\rm nt}}{v_0^{\rm tad}}
                       G^{\rm nt}(t;\tilde{v}^{\rm nt}, v_0^{\rm nt})
              \label{eq:two-pt-tad}
\end{eqnarray}
In addition, the tadpole-improvement of a simulation requires adjusting the
velocity (analogous to adjusting $\kappa$) according to $\tilde{v}^{\rm tad} =
u_0 \tilde{v}^{\rm nt}$, subject to the normalization $(v^{\rm tad})^2 = 1$ and
$(v^{\rm nt})^2 = 1$.  The adjustment on the velocity is then
\begin{equation}
\begin{array}{rcl}
v^{\rm tad}_0
& = & v^{\rm nt}_0 [1+(1-u_0^2) (v^{\rm nt}_j)^2]^{-{}^{1}\!\!/\!_{2}} \\
v^{\rm tad}_j
& = & u_0^{} v^{\rm nt}_j [1+(1-u_0^2)(v^{\rm nt}_j)^2]^{-{}^{1}\!\!/\!_{2}}.
\end{array}
\label{eq:vel-adj}
\end{equation}
The tadpole improved data is at a velocity which is shifted from the original
tadpole unimproved data.  Previous HQET calculations have either not included
tadpole improvement~\cite{Aglietti94} or have had difficulties with
it~\cite{Mandula:1997hb}.  Although one should start with a tadpole improved
action, we find it convenient to be able to tadpole improve a calculation {\it
a posteriori} because there are choices for how one can determine the
mean-field value $u_0$~\cite{Lepage93}.


\section{Appendix: Reduced Renormalizations}
\label{a:reduced}

One can define a ``reduced'' wavefunction renormalization and relate it to the
fit-model exponential.  We begin by considering the propagator as a function of
time $t$ and the residual momentum $\vec{k}$,\footnote{Recall that the residual
momentum, rather than the full momentum, is conjugate to the position.}
\begin{eqnarray}
iH(t,\vec{k})
& = & \int \frac{dk_4}{2\pi}
      \frac{e^{ik_4}}
           {\left[ v_0 \sigma \left( \frac{1}{u_0} e^{i\sigma k_4}
                                   - 1 \right)
                 + \sum_j \frac{v_j}{u_0} \sin(k_j) \right]}
      \nonumber \\
& = & \Theta\!\left(t+{}^{\underline{1-\sigma}}_{\ \;2}\right)
      \frac{u_0^{\sigma t}}{v_0}
      e^{-(t-\sigma) \ln \left( 1
                              - \sigma
                                \sum_j \frac{\tilde{v}_j}{u_0} \sin(k_j)
                                \right)^{-\sigma}}.
      \label{eq:theta}
\end{eqnarray}
Since $iH\sim e^{-\varepsilon t}$, the energy-momentum relation can be found:
\begin{eqnarray} \label{eq:e-mom}
\varepsilon
& = &-\sigma\ln\left[1-\sigma\sum_j\frac{\tilde{v}_j}{u_0}\sin(k_j)\right]
\ \approx \ \sum_j\frac{\tilde{v}_j}{u_0}\sin(k_j)
\end{eqnarray}
Aglietti~\cite{Aglietti94} notes that the energy goes to zero for both $\vec{k}
= \vec{0}$ and $\vec{k} = \vec{\pi}$, but provides a physical argument for why
this doubling problem has a negligible effect in the HQET\@.

In Eq.~(\ref{eq:theta}), it may be noticed that the $\Theta$-function has a
different argument for the different actions.  Though it was phrased
differently, this was also noticed by Davies and Thacker~\cite{Davies92} who
give recursive expressions for the Green function evolution equation for the
two cases of a forward or a backward difference in their NRQCD action.

In order to consider the renormalization effects of the fitting form, consider
the next loop-order of the propagator as a function of the time and the
residual 3-momentum,
\begin{eqnarray}
iH^{(2)}(t,\vec{k})
& = & \int \frac{dk_4}{2\pi} e^{ik_4}
      \left\{ iH(k_4,\vec{k})
            + iH(k_4,\vec{k}) \; \Sigma(k) \; iH(k_4,\vec{k}) \right\}.
      \label{eq:prop;nlo}
\end{eqnarray}
Following Aglietti~\cite{Aglietti94}, we will make use of
\begin{eqnarray}
\Sigma(k)
& = & \Sigma(0) + k_4 X_4 + \sum_j k_j X_j + O\!( k^2 ) \nonumber \\
& = & -\delta M^{\rm tad}
    + \left[ -\sigma v_0 (1-\delta Z) + \delta v_0 \right] \ln(u_0)
      \nonumber \\ &&
    + \delta Z \left[ v_0 \sigma
                      \left( \frac{1}{u_0} e^{i\sigma k_4} - 1 \right)
                    + \sum_j \frac{v_j}{u_0} \sin(k_j) \right]
      \nonumber \\ &&
    - \delta v_0 \sigma \left(\frac{1}{u_0} e^{i\sigma k_4} - 1\right)
    - \sum_j \delta{}^{\underline{v_j}}_{u_0} \sin(k_j)
\end{eqnarray}
where $\delta M^{\rm tad}$ is the tadpole improved mass renormalization (versus
$\delta M^{\rm nt}$ the not tadpole improved mass renormalization) defined by
\begin{equation} \label{eq:M-tad-not}
\delta M^{\rm tad}
= -\Sigma(0,\tilde{v}) - \sigma v_0 \ln u_0
= \delta M^{\rm nt} - \sigma v_0 \ln u_0.
\end{equation}
It may also be noticed that since $\ln u_0 \sim O\!( g^2 )$, the $[(v_0 \sigma
\delta Z + \delta v_0)\ln u_0]$ can be neglected as $O\!( g^4 )$.  We further
note that terms of the residual momentum, $O\!( \vec{k})$, can be
neglected.\footnote{The calculation including these terms is available from
author JC.}  (The residual momentum can be adjusted by introducing a ``residual
mass.'')  Finally, we note that the $\delta v_0$ and
$\delta{}^{\underline{v_j}}_{u_0}$ can be collected with the bare velocity in
precisely the proportion necessary to renormalize each velocity.  To solve
these integrals, one needs to put Eq.~(\ref{eq:prop;nlo}) into a form which
allows the use of
\begin{eqnarray}
\int_{-\infty}^{\infty} \frac{dx}{2\pi} \frac{e^{iax}}{(e^{ix}-1)}
& = & \Theta(a) \\
\int_{-\infty}^{\infty} \frac{dx}{2\pi} \frac{e^{iax}}{(e^{ix}-1)^2}
& = & (a-1) \Theta(a).
\end{eqnarray}
With these relationships, we find (eventually\footnote{We found that there are
two $\Theta$ terms.  One goes as $\Theta\!\left(t+{}^{\underline{1-\sigma}}_{\
\;2}\right)\equiv\theta_1$, the other as
$\Theta\!\left(t+{}^{\underline{3-\sigma}}_{\ \;2}\right)\equiv\theta_3$. We
resolved this assuming we were interested in late enough times ($t>-1$) that
$\theta_3=\theta_1=1$.})
\begin{eqnarray}
iH^{(2)}(t,\vec{k})
& = & \Theta\!\left(t+{}^{\underline{1-\sigma}}_{\ \;2}\right)
      \frac{u_0 ( 1 + \delta Z )}
           {v_0 \left( 1 + \frac{\delta v_0}{v_0} \right)}
      \exp\left\{ -(t-\sigma) \left[ M_\sigma \right] \right\}
      \left( 1 + O\!( g^2 ) + O\!( \tilde{v}^2 ) \right)
\nonumber
\end{eqnarray}
where $M_\sigma$ is an action-dependent function of the renormalizations, of
the velocity, and of the momentum; and $v_0 \left( 1 + {}^{\delta
v_0}\!\!/\!_{v_0} \right) = v^{\rm ren}_0$.  The relevant point is that, as was
said previously, for the forward difference action one should fit to a form of
$\exp\!(-M_f[t-1])$; whereas for the backward difference action one should fit
to a form of $\exp\!(-M_b[t+1])$ [i.e.\ fit to $\exp\{-M_\sigma(t-\sigma)\}$].
However, if one chooses to fit to the form $\exp\!(-Mt)$, then the coefficient
$Z=(1+\delta Z)$ gets changed to $Ze^{\sigma M_\sigma} \approx Z(1+\sigma
M_\sigma) \approx (1+\delta Z + \sigma M_\sigma)$.  To $O\!( g^2 )$, neglecting
$O\!( k )$ terms, $M_\sigma = (\delta M^{\rm tad} + \sigma v^{\rm ren}_0 \ln
u_0)/v^{\rm ren}_0 = -\Sigma(0)/v^{\rm ren}_0$ [Recall
Eq.~(\ref{eq:M-tad-not})].  So, to this order, the ``reduced'' wavefunction
renormalization is
\begin{eqnarray} \label{eq:Zreduced}
Z'
& = & Z - \sigma \Sigma(0)/v^{\rm ren}_0 \nonumber \\
& = & (1 + \delta Z - \sigma \Sigma(0)/v^{\rm ren}_0 ) \nonumber \\
& = & \left( 1 + \frac{g^2 C_F}{16\pi^2}
                 \left[ -2 \ln\!\left(\lambda^2a^2\right)
                      + e(\tilde{v}) - \sigma x(\tilde{v})/v^{\rm ren}_0
\right] \right).
\end{eqnarray}
This is also written in terms of the finite pieces
\begin{equation}
e'(\tilde{v}) \equiv e(\tilde{v}) - \sigma x(\tilde{v})/v^{\rm ren}_0.
\label{eq:eprime}
\end{equation}
The tadpole term is in $x(\tilde{v})$ [it gets canceled in the mass
renormalization of Eq.~(\ref{eq:x})] and, as noticed for the static case
in~\cite{Bernard:1994zh,Bernard94a}, the wavefunction and reduced wavefunction
renormalizations remain unaffected by tadpole improvement.

%

\section{Appendix: Notation}
\label{a:notation}

When comparing between the results of HQET, NRQCD and the static theory, the
difference in notation starts to become a factor.  Where Davies and Thacker
(NRQCD) used $A$ for $\Sigma\!(0)$, Aglietti (HQET) uses $A\!(p)$ for the
non-tadpole portion of the self-energy as well as using $A$ for a particular
grouping of terms for convenience in the calculation.  We are going to maintain
Davies' and Thacker's use of $A$ and give new names to Aglietti's $A$s.
However, since Aglietti considers the velocity-dependence of various
quantities, we will use Aglietti's notation for a variety of velocity-dependent
functions.  The velocity will be relevant for the HQET, but not for the {\sl
static\/} theory nor for the NRQCD\@.  In the HQET, the functional dependence
is on $\tilde{v}$ defined by
\begin{equation} \label{eq:u-def}
\tilde v = \sum_j \tilde v_j^2  = \sum_j \frac{v_j^2}{v_0^2}.
\end{equation}
Note that Aglietti calls this $u$.

Aglietti calls the mass renormalization $\delta M$; he also puts in a negative
sign, which we leave out.  Aglietti notes that for the HQET, this is velocity
dependent, and defines a function $x\!(u)$ which is proportional to his $\delta
M$
\begin{equation}
\left. \delta M \right|_{\rm Ag}
= - g^2 \left. A \right|_{\rm DT}
= \frac{g^2 C_F}{16\pi^2}
  \frac{\left. x\!(u) \right|_{\rm Ag}}{a}.
\end{equation}
$v_0$ does not appear in NRQCD and is 1 in the static limit.

In calculating the wavefunction renormalization, $\left.  \displaystyle
\frac{\partial\, \Sigma\!(p)}{\partial p_\mu\hfill} \right|_{p=0}$ is needed.
Mandula and Ogilvie use the notation $X_\mu$.  This is a useful notation and
does not conflict with either Davies and Thacker or Eichten and Hill.  Aglietti
names these as $\left. X \right|_{Ag} = \left. X_0 \right|_{MO}$ and $\left. Y
\right|_{Ag} = \left. X_3 \right|_{MO}$.  See Table~\ref{t:ag-vs-mo} for an
explicit comparison.  We choose to use Mandula's and Ogilvie's notation.

In the definition of the velocity renormalization, there is a further subtlety.
Mandula and Ogilvie consider $\delta \tilde{v}_i \equiv \tilde{v}_i^{\rm (ren)}
- \tilde{v}_i$, but Aglietti considers $\delta v_z \equiv v_z^{\rm (ren)}-v_z$
(with the definition $u_i=\tilde{v}_i={}^{v_i}\!\!/\!_{v_0}$).  As shown in
Table~\ref{t:ag-vs-mo}, a factor of $v_0^2$ must be included to translate
between $\frac{\delta v}{v}$ and $\frac{\delta \tilde{v}}{\tilde{v}}$.  In
addition, Mandula and Ogilvie include the pre-factor $\frac{g^2 C_F}{16\pi^2}$
in their definition of $c(\tilde{v})$ in Eq.~(\ref{eq:ag-deltav}).
\begin{table}[tbp]
\caption[Comparison of notation between Aglietti and Mandula and
         Ogilvie.]
        {\label{t:ag-vs-mo}
         Comparison of notation between
         Aglietti~\protect{\cite{Aglietti94}} and Mandula and
         Ogilvie~\protect{\cite{Mandula:1997hb}}.  Note also that
         Aglietti only considers motion in the $z$-direction.  Finally
         note that in the last row, Mandula and Ogilvie consider
         $\delta \tilde{v}_i$, but Aglietti considers $\delta v_z$
         ($u_i=\tilde{v}_i={}^{v_i}\!\!/\!_{v_0}$).  To convert between
         the two, one must include a factor of $v_0^2$.}
\begin{center}
\begin{tabular}{ccc} \hline \hline
Mandula and Ogilvie & Aglietti & Comparison \\ \hline
$X_0 = -i X_4$ & $X$
& $\left. X \right|_{Ag} = \left.  X_4 \right|_{MO}
                         = \left. iX_0 \right|_{MO}$ \\
$X_i$ & $Y$ & $\left. Y \right|_{Ag} = \left.  X_3 \right|_{MO}$ \\
$\displaystyle \tilde{v}_i = \frac{v_i}{v_0}$ & $\displaystyle u_z =
\frac{v_z}{v_0}$
& $\left. u_z \right|_{Ag} = \left.  \tilde{v}_z \right|_{MO}$  \\
$\begin{array}{rcl}
 \delta \tilde{v}_i
 & = & -\frac{1}{v_0} \left( X_i - \tilde{v}_i X_0 \right) \\
 & = & -\frac{1}{v_0} \left( X_i + i \tilde{v}_i X_4 \right)
 \end{array}$
&
$\begin{array}{rcl}
 \delta v_z
 & = & -i v_0 v_z X - v_0^2 Y \\
 & = & -v_0^2 \left( Y + i u_z X \right)
 \end{array}$
& $\displaystyle \frac{\delta v}{v} = v_0^2 \frac{\delta \tilde{v}}{\tilde{v}}$
\\ \hline \hline
\end{tabular}
\end{center}
\end{table}

Mandula and Ogilvie do not calculate the wavefunction renormalization,
therefore we will compare Aglietti's wavefunction renormalization to Davies and
Thacker (while using Mandula's and Ogilvie's notation for $X_\mu$).  Aglietti
uses $\delta Z = Z-1$ for the wavefunction renormalization.  To relate this to
Davies and Thacker, we note that
\begin{eqnarray}
\left. Z \right|_{Ag}
& = & 1 + \left( v_0 X_0 - \sum_j v_j X_j \right) \ = \ 1 + \frac{g^2
C_F}{16\pi^2}
          \left[ -2 \ln\!\left(a\lambda\right)^2 + e(\tilde{v}) \right]
\end{eqnarray}
where the $\ln\!\left(a\lambda\right)$ term comes from doing the self-energy
integral.  It is $\delta Z = Z-1$ which is Davies' and Thacker's $C$.
\begin{equation} \label{eq:dt;C}
\frac{C_F}{16\pi^2} \left[ -2 \ln\!\left(a\lambda\right)^2 + e(\tilde{v})
\right]
 = \left. C \right|_{DT}
 = \left. Z \right|_{DT} + \left. aA \right|_{DT}
\end{equation}
In addition, because of some discrepancies discussed in
Sec.~\ref{ss:reduced}, it will be convenient to define a ``reduced
value of $e$,'' [$e^{(R)}(\tilde{v}) \equiv e'(\tilde{v})$]:
\begin{equation}
\frac{C_F}{16\pi^2} \left[ -2 \ln\!\left(a\lambda\right)^2 + e'(\tilde{v})
\right]
 = \left. Z \right|_{DT}
\end{equation}
This reduced value can be found from $e(\tilde{v})$ and $x(\tilde{v})$ as
expressed in Eq.~(\ref{eq:eprime}).




%

\end{document}